\documentclass[twoside]{IEEEtran}

\usepackage{epsfig}
\usepackage{times}
\usepackage{float}
\usepackage{afterpage}
\usepackage{amsmath}
\usepackage{amstext}
\usepackage{amssymb,bm}
\usepackage{latexsym}
\usepackage{color}
\usepackage{graphicx}
\usepackage{amsmath}
\usepackage{amsthm}
\usepackage{graphicx}
\usepackage[center]{caption}
\usepackage{pstricks}
\usepackage{subfigure}
\usepackage{booktabs}

\newtheorem{theorem}{Theorem}
\newtheorem{remark}{Remark}

\newtheorem{proposition}{Proposition}
\newtheorem{definition}{Definition}

\newcommand{\gaplda}{3}
\newcommand{\gappdf}{1}
\newcommand{\gapnnc}{1.61}

\begin{document}

\title{On Gaussian Half-Duplex Relay Networks}

\author{Martina~Cardone, Daniela~Tuninetti, Raymond~Knopp and Umer~Salim %
\thanks{M. Cardone and R. Knopp are with the Mobile Communications Department at Eurecom, Sophia Antipolis, 06560, France (e-mail: cardone@eurecom.fr; knopp@eurecom.fr). D. Tuninetti is with the Electrical and Computer Engineering Department of the University of Illinois at Chicago, Chicago, IL 60607 USA (e-mail: danielat@uic.edu). U. Salim is with Algorithm Design group of Intel Mobile Communications, Sophia Antipolis, 06560, France (e-mail: umer.salim@intel.com).

The work of D.~Tuninetti was partially funded by NSF under award number 0643954;
the contents of this article are solely the responsibility of the author and
do not necessarily represent the official views of the NSF.
The work of D.~Tuninetti was possible thanks to the generous support of Telecom-ParisTech, Paris France,
while the author was on a sabbatical leave at the same institution.
Eurecom's research is partially supported by its industrial partners:
BMW, Cisco Systems, Monaco Telecom, Orange, SFR, ST Ericsson, SAP, Swisscom and Symantech.
The research work carried out at Intel by U. Salim has received funding from the
European Community's Seventh Framework Program (FP7/2007-2013) SACRA project (grant agreement number 249060)

The results in this paper have been submitted in part to the 2013 IEEE International Conference on Communications (ICC) and to the 2013 IEEE International Symposium on Information Theory (ISIT).}
}
\maketitle
\begin{abstract}
This paper considers Gaussian relay networks where a source transmits a message to a sink terminal with the help of one or more relay nodes. The relays work in half-duplex mode, in the sense that they can not transmit and receive at the same time. 
For the case of one relay, the generalized Degrees-of-Freedom is characterized first and then it is shown that capacity can be achieved 
to within a constant gap regardless of the actual value of the channel parameters. Different achievable schemes are presented with either deterministic or random switch for the relay node. It is shown that random switch in general achieves higher rates than deterministic switch. 
For the case of $K$ relays, it is shown that the generalized Degrees-of-Freedom can be obtained by solving a linear program and that capacity can be achieved to within a constant gap of $K/2\log(4K)$. 
This gap may be further decreased by considering more structured networks such as, for example, the diamond network.
\end{abstract}

\begin{IEEEkeywords}
Relay Channel,
Generalized Degrees-of-Freedom,
Capacity to within a Constant Gap,
Inner bound,
Outer bound,
Half-duplex.
\end{IEEEkeywords}

\newcommand{\eu}{{\rm e}}
\newcommand{\jj}{{\rm j}}

\section{Introduction}
\label{sec:intro}

The performance of wireless systems can be enhanced by enabling cooperation between the wireless nodes.
The simplest form of cooperation is modeled by the Relay Channel (RC) where a source terminal communicates to a destination with the help of a relay node. In this multi-hop system the relay helps to increase the coverage and the throughput of the network.

Relays employed in practical wireless networks can be classified into two categories: Full-Duplex (FD) and Half-Duplex (HD). In the former case the relays transmit and receive simultaneously; in the latter case the relays can either transmit or receive at any given time, but not both. 
There are some relatively expensive relay devices which work in FD mode, normally used in military communications. However FD relaying in wireless networks has practical restrictions such as self-interference, which make the implementation of decoding algorithm challenging.
As a result HD relaying proves to be a more practical technology with its relatively simple signal processing. Thus it is more realistic to assume that the relay operates in HD mode either in Frequency Division Duplexing (FDD) or Time Division Duplexing (TDD). In FDD, the relay uses one frequency band to transmit and another one to receive; in TDD, the relay listens for a fraction $\gamma\in[0,1]$ of time and then transmits in the remaining  $1-\gamma$ time fraction. From an application point of view, the HD model fits future 4G network with relays \cite{3GPPRel10doc}, where the relay communicates over-the-air with the source, which is called the {\em Donor-eNB}. We keep our focus on deployment scenarios where the relay works in TDD HD mode.

In this work we concentrate on the HD relay networks, where the relays transmit and receive in different time slots. HD relaying has received considerable attention lately, as summarized next.

\subsection{Related Work}

\paragraph{Single Relay Networks}
The RC was first introduced by van der Meulen~\cite{Meulen} and then thoroughly studied by Cover and El Gamal~\cite{coveElGamal}. In~\cite{coveElGamal} the authors study the general memoryless RC, derive inner and outer bounds on the capacity and establish the capacity for some classes of RCs. The proposed outer bound is now known as the {\em  max-flow min-cut outer bound}, or cut-set for short, which can be extended to more general memoryless networks~\cite{book:ElGamalKim2012}. Two achievable relaying strategies were proposed in~\cite{coveElGamal}, whose combination is still the largest known achievable rate for a general RC, namely Decode-and-Forward (DF) and Compress-and-Forward (CF). In DF, the relay fully decodes the message sent by the source and then coherently cooperates with the source to communicate this information to the destination. In CF, the relay does not attempt to recover the source message, but it just compresses the information received and then sends it to the destination. 
The capacity of the general memoryless RC is known for some special classes, in particular for degraded RCs, reversely degraded RCs and semi-deterministic RC~\cite{coveElGamal}.

The HD-RC was studied by Host-Madsen in~\cite{Host2002}. Here the author derives both an upper and a lower bound on the capacity. The former is based on the cut-set arguments, the latter exploits the {\em Partial-Decode-and-Forward} (PDF) strategy where the relay only decodes part of the message sent by the source. Host-Madsen considers the transmit/listen state of the relay as fixed and therefore known a priori to all nodes. 

In \cite{kramer-allerton}, Kramer shows that larger rates can be achieved by using a random transmit/listen switch strategy at the relay. In this way, the source and the relay can harness the randomness that lies in the switch in order to transmit extra information. An important observation of \cite{kramer-allerton} is that there is no need to develop a separate theory for memoryless networks with HD nodes as the HD constraints  can be incorporated into the memoryless FD framework. In this work we shall adopt this approach in deriving outer and inner bounds for HD relay networks. 


\paragraph{Multiple Relay Networks}
The pioneering work of Cover and El Gamal~\cite{coveElGamal} has been extended to networks with multiple relays. In~\cite{KramerGastparGuptaIT05} Kramer {\em et al.} proposed several inner and outer bounds as a generalization of DF, CF and the cut-set bound. It was shown that DF achieves the ergodic capacity of a wireless Gaussian network with phase fading if phase information is available only locally and the relays are close to the source node. 

The exact characterization of the capacity region of a general memoryless network is challenging. Recently it has been advocated that progress can be made towards understanding the capacity by showing that achievable strategies are provably ``close'' to (easily computable) outer bounds \cite{AvestimehrThesis}. As an example, Etkin {\em et al.} characterized the capacity of the Gaussian Interference Channel to within 1 bit regardless of the system parameters~\cite{etw}.
In \cite{avestimher:netflow}, the authors study unicast and multicast Gaussian relay networks with $K$ nodes and show that capacity can be achieved to within $\sum_{k=1}^{K}5\min\{M_k,N_k\}$ bits with {\em quantize-remap-and-forward} (QMF), where $M_k$ and $N_k$ are the number of transmit and receive antennas, respectively, of node $k\in[1:K]$. Interestingly, the result is valid for static and ergodic fading networks where the nodes operate either in FD mode or in HD mode with deterministic listen/transmit schedule for the relays.
For single antenna systems, Lim {\em et al.} in~\cite{nncLim} recently showed that this $5K$~bits~gap can be reduced to $0.6 K$~bits for FD relay networks with {\em noisy network coding} (NNC). Both QMF and NNC are network extensions of CF.
 
The gap characterization of~\cite{nncLim} is valid for any multi relay network but linear in the number of nodes in the network, which could be a too coarse capacity characterization for networks with a large number of nodes. Tighter gaps can be obtained for more structured networks. For example, the {\em diamond network} model was first proposed in~\cite{ScheinGallagerISIT200}. A diamond network consists of a source, a destination and $K-2$ relays. The source and the destination cannot communicate directly and the relays cannot communicate among themselves. In other words, a general Gaussian multi relay network with $K$ nodes is characterized by $K(K-1)$ channel gains, while a diamond network only has $2(K-2)$ non zero channel gains. In~\cite{ScheinGallagerISIT200} the case of two relays was studied for which an achievable region based on time sharing between DF and {\em  amplify-and-forward} (AF) was proposed.
The capacity of a general FD diamond network is known to within $2\log(K-1)$~bits~\cite{ozgunITW2012}. 
If in addition the FD diamond network is symmetric, that is, all source-relay links are equal and all relay-destination links are equal, the gap is less than 2~bits for any $K$~\cite{NiesenDiggavi}. 
HD diamond networks have been studied as well, albeit only for deterministic switch for the relays. In a HD diamond network with $(K-2)$ relays, there are $2^{K-2}$ possible combinations of listening and transmit states, since each relay can either transmit or receive. For the case of $(K-2)=2$ relays, \cite{Bagheri2009} shows that out of $2^{K-2}=4$ possible states only $(K-1)=3$ states suffice to achieve capacity to within less than 4~bits. Their achievable scheme is a clever extension of the two-hop DF strategy of~\cite{XueSandhuIT07}. It is interesting to note that \cite{Bagheri2009} derived closed-form expressions for the fractions of time the relays are active and tight outer bounds based on the dual of the linear program (LP) associated with the classical cut-set bound. Extensions of these ideas to more than two relays appear difficult due to the combinatorial structure of the problem.

Inspired by \cite{Bagheri2009}, the authors in~\cite{Fragouli2012} showed that for a very specific HD diamond network with $(K-2)=3$ relays, $(K-1)=4$ states out of $2^{K-2}=8$ are active in the cut-set outer bound. In the same work, it was verified numerically that for a general HD diamond network with $(K-2)\leq 7$ relays, $(K-1)$ states suffice for the cut-set upper bound and it is conjectured that the same holds for any number of relays. We remark that in~\cite{Fragouli2012} only the cut-set upper bound was considered; moreover only the case of deterministic switch and per-symbol power constraint was considered. 

Multi relay networks were also studied in~\cite{OngMultiRelay} where the authors determine numerically the optimal fractions of time each relay transmits/receives with DF through an iterative algorithm. Also in this case the relays use deterministic switch.



\subsection{Contributions}
In this work we focus on the HD relay networks. The exact capacity of this channel is unknown. In this paper we make progress toward determining its capacity by giving a constant gap result for any Gaussian network with random switch. Our main contribution can be summarized as follows:
\begin{enumerate}

\item 
We determine the {\em generalized Degrees-of-Freedom} (gDoF) of the HD RC with a single relay.
We identify three schemes that achieve the gDoF upper bound.
The simplest one is inspired by the Linear Deterministic Approximation (LDA) of the Gaussian noise channel at high SNR \cite{avestimher:netflow}; it uses superposition coding at the source, DF at the relay and stripping decoding at both the relay and destination; we note that neither power allocation nor backward decoding is required at the nodes.
The second and third schemes use more sophisticated coding techniques and are based on PDF and NNC strategies~\cite{book:ElGamalKim2012}. 

\item
We prove that the three schemes above achieve the capacity to within a constant gap regardless of the channel parameters. We consider both deterministic and random switch for the relay.
Thus in the second case the relay harnesses the randomness that lies in the switch to achieve larger rates and therefore smaller gaps from the cut-set upper bound.

\item
We prove that PDF with random switch is optimal for a Gaussian diamond network with one relay, i.e., it achieves the capacity, even though we were not able to determine the capacity achieving input distribution.

\item
We determine the capacity of the noiseless LDA channel. In particular we show that random switch and non-uniform inputs at the relay are optimal.


\item
For HD networks with $K$ nodes (HD-MRC), all of which employ random switch, we prove that NNC achieves the cut-set outer bound to within $K/2\log(4K)$~bits. For diamond relay networks assuming that the conjecture in~\cite{Fragouli2012} holds for any $K$, the gap can be reduced to $5\log(K)$. 


\end{enumerate}

\subsection{Paper Organization}
The rest of the paper is organized as follows. 
Section \ref{sec:channel model} describes the channel model and states our main result. 
Section \ref{sec:cutset} derives the gDoF upper-bound based on the cut-set bound. 
Section \ref{sec:pdf inner} provides a lower bound based on the PDF strategy.
Section \ref{sec:lda} highlights a motivating example based on the LDA and provides a simple achievable based on this approach. 
Sections \ref{sec:Analytical Gaps} and \ref{sec:Numerical Gaps} are devoted to the analytical and numerical proofs that the capacity for a single relay network is achievable to within a constant gap, respectively. 
Section \ref{sec:multirelay} considers networks with multiple relays and it shows that the gDoF can be computed by solving a linear program and that NNC achieves the capacity to within a constant gap. 
Section \ref{sec:conclusion} concludes the paper.

\section{Single Relay Networks: System Model}
\label{sec:channel model}

\subsection{General Memoryless Relay Channel}
\label{sec:channel model:general}

A RC consists of two input alphabets $\left (\mathcal{X}_s,\mathcal{X}_r \right )$, two output alphabets $\left (\mathcal{Y}_{r},\mathcal{Y}_d \right )$ and a transition probability $P_{{Y}_{r},{Y}_d|{X}_s,{X}_r}$.
The source sends symbols from $\mathcal{X}_s$, the relay receives symbols in $\mathcal{Y}_r$ and sends symbols from $\mathcal{X}_r$, and the destination receives symbols in $\mathcal{Y}_d$.
The source has a message $W\in[1:2^{N R}]$ for the destination where $N$ denotes the codeword length and $R$ the transmission rate in bits per channel use \footnote{Logarithms are in base 2.}. At time $i$, $i \in [1:N]$, the source maps its message $W$ into a channel input symbol $X_{s,i}(W)$ and the relay maps its past channel observations into a channel input symbol $X_{r,i}(Y_{r}^{i-1})$.
The channel is assumed to be memoryless, that is, the following Markov chain holds for all $i\in[1:N]$
\begin{align*}
(W,Y_{r}^{i-1},Y_d^{i-1},X_s^{i-1},X_{r}^{i-1}) \to (X_{s,i},X_{r,i}) \to (Y_{r,i},Y_{d,i}).
\end{align*}
At time $N$, the destination makes an estimate $\widehat{W}(Y_d^N)$ of the message $W$ based on all its channel observations $Y_d^N$. 
A rate $R$ is said to be $\epsilon$-achievable if $\mathbb{P}[ \widehat{W} \neq W ] \leq \epsilon$ for some $\epsilon\in[0,1]$. The capacity is the largest nonnegative rate  that is $\epsilon$-achievable for any $\epsilon>0$.

We note that half-duplex channels are a special case of the memoryless full-duplex framework in the following sense~\cite{kramer-allerton}: let the channel input of the relay be the pair $(X_r,S_r)$, where $X_r\in \mathcal{X}_r$ as before and $S_r\in\{0,1\}$ is the {\em state} random variable that indicates whether the relay is in receive-mode ($S_r=0$) or in transmit-mode ($S_r=1$). The memoryless channel transition probability is defined as
\begin{align*}
P_{{Y}_{r},{Y}_{d}|{X}_s,{X}_r,S_r=0} 
 &= P^{(0)}_{{Y}_{r},{Y}_{d}|{X}_s,S_r=0}
\\
P_{{Y}_{r},{Y}_{d}|{X}_s,{X}_r,S_r=1} 
 &= P^{(1)}_{{Y}_d|{X}_s,{X}_r,S_r=1} P^{(1)}_{{Y}_{r}|S_r=1},
\end{align*}
that is, when the relay is in receive-mode ($S_r=0$) the outputs ${Y}_{r},{Y}_{d}$ are independent of $X_r$ and when the relay is in transmit-mode ($S_r=1$) the relay output ${Y}_{r}$ is independent of everything else. In other words, the (still memoryless) channel is now specified by the two transition probabilities one for each mode of operation~\cite{kramer-allerton}.

\subsection{The Gaussian Half-Duplex RC}
\label{sec:channel model:G-HD-RC model}

We consider a single-antenna complex-valued Gaussian Half-Duplex Relay Channel (G-HD-RC), shown in Fig.~\ref{fig:fig1}, where the inputs are subject to an average power constraint, described by the input/output relationship
\begin{subequations}
\begin{align}
Y_{r} &= \sqrt{C} X_s \ (1-S_r)                               + Z_r, \\
Y_{d} &= \sqrt{S} X_s + \eu^{\jj \theta} \sqrt{I}  X_r \ S_r  + Z_d, \ \theta\in\mathbb{R},
\end{align}
\label{eq:awgn half}
\end{subequations}
where the channel gains $C,S,I$ are constant and therefore known to all terminals. Without loss of generality we can assume that the relay output $Y_{r}$ does not contain the relay input $X_r$ because the relay node can always subtract $X_r$ from $Y_{r}$. Moreover, since a node can compensate for the phase of one of its channel gains, we can assume without loss of generality that the channel gains from the source to the other two terminals are real-valued and nonnegative. 
The channel inputs are subject to unitary average power constraints without loss of generality, i.e., $\mathbb{E}[ |X_u|^2 ] \leq 1, \ u\in\{s,r\}$. The `switch' random variable $S_r$ is binary.
The noises $Z_d,Z_r$ are assumed to be zero-mean jointly Gaussian and with unit power without loss of generality. In particular (but not without loss of generality) in this work we assume that $Z_d$ is independent of $Z_r$.
In the following we will only consider G-HD-RC for which $C>0$ and $I>0$ in~\eqref{eq:param high snr}, since for either $C=0$ or $I=0$ the relay is disconnected from either the source or the destination, respectively, so the channel reduces to a point-to-point channel with capacity $\log(1+S)$.

The capacity of the channel in~\eqref{eq:awgn half} is unknown. Here we make progress toward determining its capacity by establishing its gDoF, i.e., an exact capacity characterization in the limit for infinite SNR~\cite{etw}, and its capacity to within a constant gap at any finite SNR. Consider $\mathsf{SNR}>0$ and the parameterization
\begin{subequations}
\begin{align}
   S &:= \mathsf{SNR}^{\beta_{\rm sd}}, \text{source-destination link},
\\ I &:= \mathsf{SNR}^{\beta_{\rm rd}}, \text{relay-destination link},
\\ C &:= \mathsf{SNR}^{\beta_{\rm sr}}, \text{source-relay link},
\end{align}
\label{eq:param high snr}
\end{subequations}
for some $(\beta_{\rm sd},\beta_{\rm rd},\beta_{\rm sr})\in\mathbb{R}_{+}^3$.
We define:
\begin{definition}
The gDoF is 
\begin{align*}
\mathsf{d}^{\rm(HD-RC)} &:= \lim_{\mathsf{SNR}\to+\infty} \frac{C^{\rm(HD-RC)}}{\log(1+\mathsf{SNR})},
\end{align*}
where $C^{\rm(HD-RC)}$ is the capacity of the G-HD-RC.
\end{definition}
\begin{definition}
The capacity $C^{\rm(HD-RC)}$ is said to be known to within $\mathsf{b}$~bits if one can show rates $R^{\rm(in)}$ and $R^{\rm(out)}$ such that
\[
R^{\rm(in)}\leq C^{\rm(HD-RC)}\leq R^{\rm(out)}
\leq R^{\rm(in)} + \mathsf{b}\log(2).
\]
\end{definition}

Our main result for single relay networks can be summarized as
\begin{theorem}
\label{thm:main 1 relay}
The gDoF of the G-HD-RC is given by \eqref{eq:dofHDrelay} at the top of next page
\begin{figure*}
\begin{align}
\mathsf{d}^{\rm(HD-RC)}
=
\left\{\begin{array}{ll}
\beta_{\rm sd}+\frac{(\beta_{\rm rd}-\beta_{\rm sd})(\beta_{\rm sr}-\beta_{\rm sd})}{(\beta_{\rm rd}-\beta_{\rm sd})+(\beta_{\rm sr}-\beta_{\rm sd})} & \text{for} 
\ \beta_{\rm sr}> \beta_{\rm sd}, 
\ \beta_{\rm rd}> \beta_{\rm sd} \\
\beta_{\rm sd}        & \text{otherwise}. \\
\end{array}\right.
\label{eq:dofHDrelay}
\end{align}
\end{figure*}
and the cut-set upper bound is achieved to within the following number of bits
\begin{center}
\begin{tabular}{|l|l|l|l|}
\hline
Achievable scheme  & \multicolumn{1}{|c|}{LDA} & \multicolumn{1}{|c|}{NNC} &  \multicolumn{1}{|c|}{PDF} \\  
\hline
analytical gap & \gaplda & \gapnnc & \gappdf \\
numerical  gap & 1.59    &  1.52   & 1 \\
\hline
\end{tabular}
\end{center}
where LDA is a very simple achievable scheme inspired by the linear deterministic approximation of the G-HD-RC at high SNR,
PDF is partial-decode-and-forward and NNC is noisy-network-coding, or compress-and-forward.
\end{theorem}
Sections~\ref{sec:cutset}-\ref{sec:Numerical Gaps} are devoted to the proof of Theorem~\ref{thm:main 1 relay}.

\begin{remark}
The gDoF of the Gaussian Full-Duplex Relay Channel (G-FD-RC) is
\begin{align}
\mathsf{d}^{\rm(FD-RC)}
=\beta_{\rm sd} 
+ \min\{[\beta_{\rm sr}-\beta_{\rm sd}]^+,[\beta_{\rm rd}-\beta_{\rm sd}]^+\},
\label{eq:dofFDrelay}
\end{align}
and its capacity $C^{\rm(FD-RC)}$ is achievable to within $1$~bit per channel use~\cite{avestimher:netflow}. 
We notice that HD achieves the same gDoF of FD if $\min\{\beta_{\rm rd},\beta_{\rm sr}\}\leq \beta_{\rm sd}$, in which case the RC behaves gDoF-wise like a point-to-point channel from TX to RX with gDoF given by $\beta_{\rm sd}$.
In either FD or HD the gDoF has a `routing' interpretation~\cite{avestimher:netflow}: if the weakest link from the source to the destination through the relay is smaller than the direct link from the source to the destination then direct transmission is optimal and the relay can be kept silent, otherwise it is optimal to communicate with the help of the relay.
\end{remark}

\section{Single Relay Networks: Upper Bound}
\label{sec:cutset}
This section is devoted to the proof of a number of upper bounds that we shall use for the converse part of Theorem~\ref{thm:main 1 relay}. 
From the cut-set bound we have:
\begin{proposition}
\label{prop:cutset upper bounds}
The capacity of the G-HD-RC is upper bounded as in~\eqref{eq:cuset for gap pdf random},~\eqref{eq:cuset for gap pdf fixed and numerical} and~\eqref{eq:cuset for gap analytical} at the top of next page
\begin{figure*}
\begin{align}
C^{\rm(HD-RC)}&\leq  \min
\left.\Big\{I(X_s,X_r,S_r;Y_d),I(X_s;Y_r,Y_d|X_r,S_r)\Big\}\right|_{(X_s,X_r,S_r) \sim P^*_{X_s,X_r,S_r}} 
\label{eq:cuset for gap pdf random}
\\&\leq  \max \min 
\Big\{
\mathcal{H}(\gamma) + \gamma I_{1} + (1-\gamma) I_{2},
                      \gamma I_{3} + (1-\gamma) I_{4}
\Big\}=: r^{\rm(CS-HD)}
\label{eq:cuset for gap pdf fixed and numerical}
\\&\leq  
2\log(2)+ \log \left (1+S \right )\left(1+\frac{(b_1 -1)(b_2-1)}{(b_1-1) + (b_2-1)}\right),
\label{eq:cuset for gap analytical}
\end{align}

\begin{align}
C^{\rm(HD-RC)} &\geq  \min
\Big\{I(X_s,X_r,S_r; Y_d), \nonumber\\&
\left.  I(U; Y_r|X_r,S_r)+I(X_s; Y_d|X_r,S_r,U)\Big\}\right|_{(X_s,X_r,S_r) \sim P^*_{X_s,X_r,S_r} \ \text{\rm and $U=X_r$ or $U=X_rS_r+X_s(1-S_r)$}},
\label{eq:inner for gap pdf random}
\\
C^{\rm(HD-RC)} &\geq  \max \min 
\Big\{
I_0^{\rm(PDF)} + \gamma I_{5} + (1-\gamma) I_{6},
                 \gamma I_{7} + (1-\gamma) I_{8}
\Big\}=: r^{\rm(PDF-HD)}
\label{eq:inner for gap pdf fixed and numerical}
\\&\geq   
\log \left (1+S \right )\left(1+\frac{(c_1 -1)(c_2-1)}{(c_1-1) + (c_2-1)}\right),
\label{eq:inner for gap analytical}
\end{align}

\begin{align}
r^{\rm(LDA-HD)}
\!:=\!  \log(1+S)
\!+\! \frac{ \log\left(1+\frac{I}{1+S}\right) }
       { \log\left(1\!+\!\frac{I}{1+S}\right)\!+\!\left [\log\left(1\!+\!\frac{C}{1\!+\!S}\right) \!-\!\log\left(1\!+\!\frac{S}{1\!+\!S}\right) \right ]^+} 
\left[  \log\left(1\!+\!\frac{C}{1+S}\right) \!-\! \log\left( 1\!+\! \frac{S}{1\!+\!S} \right) \right]^+.
\label{eq:lda inner bound}
\end{align}
\end{figure*}
where
\begin{itemize}

\item
in~\eqref{eq:cuset for gap pdf random}: the distribution $P^*_{X_s,X_r,S_r}$ is the one that maximizes the cut-set upper bound,

\item
in~\eqref{eq:cuset for gap pdf fixed and numerical}:
the parameter $\gamma:=\mathbb{P}[S_r=0]\in[0,1]$ represents the fraction of time the relay node listens, 
$\mathcal{H}(\gamma)$ is the binary entropy function defined as
\begin{align}
\mathcal{H}(\gamma):= -\gamma\log(\gamma)-(1-\gamma)\log(1-\gamma),
\label{eq:binaty entropy function}
\end{align}
the maximization is over the set
\begin{align}
  & \gamma\in[0,1], \
\label{eq:GHDRC CS gamma}
\\& |\alpha_1| \leq 1, \ 
\label{eq:GHDRC CS alpha1}
\\& (P_{s,0},P_{s,1},P_{r,0},P_{r,1}) \in \mathbb{R}^4_{+} \nonumber
\\& : \gamma P_{u,0} + (1-\gamma) P_{u,1} \leq 1, \ u\in\{s,r\},
\label{eq:GHDRC CS powers}
\end{align}
and the mutual informations $I_{1}, \ldots, I_{4}$ are defined as
\begin{align}
   I_{1} &:= \log\left(1+ S \ P_{s,0}\right),
\label{eq:def of I1}
\\ I_{2} &:= \log\left(1\!+\! S \! P_{s,1}\!+\! I \! P_{r,1} \!+\!2|\alpha_1|\sqrt{  S \! P_{s,1}  I \! P_{r,1}}\right),
\label{eq:def of I2}
\\ I_{3} &:= \log\left(1+(C +S) P_{s,0}\right),
\label{eq:def of I3}
\\ I_{4} &:= \log\left(1+(1-|\alpha_1|^2) S \ P_{s,1}\right),
\label{eq:def of I4}
\end{align}

\item in~\eqref{eq:cuset for gap analytical}: the terms $b_1$ and $b_2$ are defined as
\begin{align}
  & b_1 := \frac{\log \left(1+(\sqrt{I}+\sqrt{S})^2\right)}{\log \left (1+S \right )} > 1 \ \text{since $I>0$},
\label{eq:def of b1}
\\& b_2 := \frac{\log \left (1+C+S \right)}{\log \left (1+S \right )} > 1 \ \text{since $C>0$}.
\label{eq:def of b2}
\end{align}

\end{itemize}
\end{proposition}
\begin{IEEEproof}
The proof and the definitions of the above quantities can be found in Appendix~\ref{app: proof of prop:cutset upper bounds}.
\end{IEEEproof}
The upper bound in~\eqref{eq:cuset for gap pdf random} will be used to prove that PDF with random switch achieves capacity to within~\gappdf~bit, the one in~\eqref{eq:cuset for gap pdf fixed and numerical} to prove that PDF with deterministic switch also achieves capacity to within~\gappdf~bit and for numerical evaluations (since we do not know the distribution $P^*_{X_s,X_r,S_r}$ that maximizes the cut-set upper bound in~\eqref{eq:cuset for gap pdf random}), and the one in~\eqref{eq:cuset for gap analytical} for analytical computations such as the derivation of the gDoF.

\begin{proposition}
\label{prop:cutset dof}
The gDoF of the G-HD-RC is upper bounded by the right hand side of~\eqref{eq:dofHDrelay}.
\end{proposition}
\begin{IEEEproof}
The proof can be found in Appendix~\ref{app: proof of prop:cutset dof}.
\end{IEEEproof}

\section{Single Relay Networks: Lower Bounds based on partial-decode-and-forward}
\label{sec:pdf inner}
This section is devoted to the proof of a number of lower bounds that we shall use for the direct part of Theorem~\ref{thm:main 1 relay}. 
From the achievable rate with PDF we have:
\begin{proposition}
\label{prop:inner upper bounds}
The capacity of the G-HD-RC is lower bounded as in~\eqref{eq:inner for gap pdf random},~\eqref{eq:inner for gap pdf fixed and numerical} and~\eqref{eq:inner for gap analytical} at the top of next page where
\begin{itemize}

\item
in~\eqref{eq:inner for gap pdf random}: we fix the input $P_{U,X_s,X_r,S_r}$ to evaluate the PDF lower bound; in particular we set $P_{X_s,X_r,S_r}$ to be the same distribution that maximizes the cut-set upper bound in~\eqref{eq:cuset for gap pdf random} and we choose either $U=X_r$ or $U=X_rS_r+X_s(1-S_r)$.

\item
in~\eqref{eq:inner for gap pdf fixed and numerical}: the parameter $\gamma:=\mathbb{P}[S_r=0]\in[0,1]$ represents the fraction of time the relay node listens, the maximization is over the set~\eqref{eq:GHDRC CS gamma}-\eqref{eq:GHDRC CS powers} as for the cut-set upper bound in~\eqref{eq:cuset for gap pdf fixed and numerical}, the mutual informations  $I_{5}, \ldots, I_{8}$ are 
\begin{align}
   I_{5} &:= I_{1} \ \text{\rm in~\eqref{eq:def of I1}},
\label{eq:def of I5}
\\ I_{6} &:= I_{2} \ \text{\rm in~\eqref{eq:def of I2}},
\label{eq:def of I6}
\\ I_{7} &:= \log\left(1+\max\{C,S\} P_{s,0}\right) \leq I_{3} \ \text{\rm in~\eqref{eq:def of I3}},
\label{eq:def of I7}
\\ I_{8} &:= I_{4} \ \text{\rm in~\eqref{eq:def of I4}},
\label{eq:def of I8}
\end{align}
and $I_0^{\rm(PDF)} := I(S_r; Y_d)$ is computed from the density
\begin{align}
 f_{Y_d}(t) 
 &\!=\! \frac{  \gamma}{\pi v_0} \! \exp(-|t|^2/v_0)
  \!+ \!\frac{1-\gamma}{\pi v_1}\! \exp(-|t|^2/v_1),
\end{align}
with $t\in\mathbb{C}, v_0 = \exp(I_5), v_1 = \exp(I_6)$.

\item
in~\eqref{eq:inner for gap analytical}: the terms $c_1$ and $c_2$ are
\begin{align}
  & c_1 := \frac{\log \left(1+I+S\right)}{\log \left (1+S \right )} > 1 \ \text{since $I>0$},
\label{eq:def of c1}
\\& c_2 := \frac{\log \left (1+\max\{C,S\} \right)}{\log \left (1+S \right )} > 1 \ \text{since $C>0$}.
\label{eq:def of c2}
\end{align}

\end{itemize}
\end{proposition}
\begin{IEEEproof} 
The proof can be found in Appendix~\ref{app: proof of prop:inner upper bounds}.
\end{IEEEproof} 
The lower bound in~\eqref{eq:inner for gap pdf random} will be compared to the upper bound in~\eqref{eq:cuset for gap pdf random} to prove that PDF with random switch achieves capacity to within~\gappdf~bit, the one in~\eqref{eq:inner for gap pdf fixed and numerical} with the one in~\eqref{eq:cuset for gap pdf fixed and numerical} to prove that PDF with deterministic switch also achieves capacity to within~\gappdf~bit and for numerical evaluations, and the one in~\eqref{eq:inner for gap analytical} for analytical computations such as evaluation of the achievable gDoF.

\begin{proposition}
\label{prop:inner dof}
The gDoF of the G-HD-RC is lower bounded by the right hand side of~\eqref{eq:dofHDrelay}.
\end{proposition}
\begin{IEEEproof}
The proof can be found in Appendix~\ref{app: proof of prop:inner dof}.
\end{IEEEproof}
Propositions~\ref{prop:cutset dof} and~\ref{prop:inner dof} show that the gDoF for the G-HD-RC is given by~\eqref{eq:dofHDrelay}.

\section{Single Relay Networks: A Simple Achievable Strategy}
\label{sec:lda}
In this section we propose a very simple achievable scheme that is gDoF optimal, that achieves capacity to within \gaplda~bits and that can be implemented in practical HD relay networks.
In Section~\ref{sec:achievLDA} we describe a deterministic-switch achievable strategy for the Linear Deterministic Approximation (LDA) of the G-HD-RC at high SNR which we mimic in Section~\ref{sec:ach simple} to derive an achievable rate for the G-HD-RC at any SNR. This achievable scheme is referred to as the LDA-strategy, or LDA for short.
The main result of this section is:
\begin{proposition}
\label{prop:lda1}
The capacity of the G-HD-RC is lower bounded as in~\eqref{eq:lda inner bound} at the top of this page.
\end{proposition}
The rest of the section is devoted to the proof of Proposition~\ref{prop:lda1}.
Before we provide the details of the scheme, we point out three important practical aspects of this scheme that are worth noticing:
\begin{enumerate}
\item the destination does not use backward decoding, which simplifies the decoding procedure and incurs no delay, 
\item the destination uses successive decoding, which is simpler than joint decoding, and
\item no power allocation is applied at the source or at the relay, which simplifies the encoding procedure and can be used for time-varying channel as well. The source uses superposition coding to `route' part of its data through the relay.
\end{enumerate}
These aspects will be clear from the actual description of the scheme.
Moreover we can show that
\begin{proposition}
\label{prop:lda}
The LDA strategy achieves the gDoF upper bound in~\eqref{eq:dofHDrelay}.
\end{proposition}
\begin{IEEEproof}
The proof can be found in Appendix~\ref{app: proof of prop:inner dof LDA}.
\end{IEEEproof}

\subsection{A Motivating Example}
\label{sec:achievLDA}
The LDA of the G-HD-RC in~\eqref{eq:awgn half} is a deterministic channel with input-output relationship
\begin{subequations}
\begin{align}
Y_{r} &= \mathbf{S}^{n-\beta_{\rm sr}} X_s \ (1-S_r), \\
Y_{d} &= \mathbf{S}^{n-\beta_{\rm sd}} X_s + \mathbf{S}^{n-\beta_{\rm rd}} X_r \ S_r,
\end{align}
\label{eq:chmodel GHDRC LDA}
\end{subequations}
for some nonnegative integers $\beta_{\rm sr},\beta_{\rm sd},\beta_{\rm rd}$,
where the inputs and outputs are vectors of length $n:=\max\{\beta_{\rm sr},\beta_{\rm sd},\beta_{\rm rd}\}$ and $\mathbf{S}$ is the $n\times n$ shift matrix~\cite{avestimher:netflow}. 

The capacity of a deterministic RC is given by the cut-set upper bound~\cite{avestimher:netflow}.
For the LDA in~\eqref{eq:chmodel GHDRC LDA} the cut-set upper-bound evaluates to
\begin{theorem}
\label{prop:capacity of the deterministic LDA}
The capacity of the deterministic HD RC in~\eqref{eq:chmodel GHDRC LDA} is given by~\eqref{eq:exact capacity of LDA HD RC} at the top of this page
\begin{figure*}
\begin{align}
C^{\rm(HD)}
= \left\{\begin{array}{ll}
\beta_{\rm sd} + 
  \max_{\gamma\in[0,1]}\min\Big\{ 
       (1-\theta^*\left( \gamma\right))\log\frac{1}{1-\theta^*\left( \gamma\right)} +\theta^*\left( \gamma\right) \log\frac{L-1}{\theta^*\left( \gamma\right)},
       \gamma [\beta_{\rm sr}-\beta_{\rm sd}]^+
\Big\} & \text{for} 
\ \beta_{\rm sr}> \beta_{\rm sd}, 
\ \beta_{\rm rd}> \beta_{\rm sd} \\
\beta_{\rm sd}        & \text{otherwise}.
\end{array}\right.
\label{eq:exact capacity of LDA HD RC} 
\end{align}

\begin{align}
  \max\{R\} 
           &=\max_{P_{X_s,X_r,S_r}} \min \Big\{I(X_s,X_r,S_r;Y_d),I(X_s;Y_r,Y_d|X_r,S_r)\Big\}
\nonumber\\&=\max_{P_{X_s,X_r,S_r}} \min \Big\{H(Y_d),H(Y_r,Y_d|X_r,S_r)\Big\}
\nonumber\\&\leq \max_{P_{X_s,X_r,S_r}} \min \Big\{H(Y_d|S_r),H(Y_r,Y_d|X_r,S_r)\Big\}+H(S_r)
\nonumber\\
&\leq \max_{\gamma\in[0,1]}\min\{
     \gamma  \beta_{\rm sd} 
+ (1-\gamma) \max\{\beta_{\rm sd},\beta_{\rm rd}\},
      \gamma  \max\{\beta_{\rm sd},\beta_{\rm sr}\}
+ (1-\gamma)  \beta_{\rm sd}
\}+\log(2)
\nonumber\\&=
\beta_{\rm sd} +   \gamma^{*}_{\rm LDA} [\beta_{\rm sr}-\beta_{\rm sd}]^+ +\log(2),
\label{eq:cs lda}
\end{align}
\end{figure*}
where $\theta^*\left( \gamma\right) = 1-\max\{1/L,\gamma\}$ and $L := 2^{[\beta_{\rm rd}- \beta_{\rm sd}]^+}$.
\end{theorem}
\begin{IEEEproof}
The proof can be found in Appendix~\ref{app:exact capacity of LDA HD RC}. 
\end{IEEEproof}

Next we further upper bound the capacity in~\eqref{eq:exact capacity of LDA HD RC} because our goal is to get insights into asymptotically optimal strategies for the G-HD-RC. For the channel in~\eqref{eq:chmodel GHDRC LDA} we have that~\eqref{eq:cs lda} at the top of this page holds, where $\gamma^{*}_{\rm LDA}$ is the optimal $\gamma:=\mathbb{P}[S_r=0]\in[0,1]$ obtained by equating the two arguments within the $\min$ and is given by
\begin{align*}
\gamma^{*}_{\rm LDA} := 
\left\{\begin{array}{ll}
\frac{(\beta_{\rm rd}-\beta_{\rm sd})}{(\beta_{\rm rd}-\beta_{\rm sd})+(\beta_{\rm sr}-\beta_{\rm sd})} & \text{for} \ \beta_{\rm rd}> \beta_{\rm sd}, \ \beta_{\rm sr}> \beta_{\rm sd} \\
0        & \text{otherwise}. \\
\end{array}\right.
\end{align*}

Next we show that the upper bound in~\eqref{eq:cs lda} is achievable to within $\log(2)=1$~bit. This 1~bit represents the maximum amount of information $I(S_r;Y_d)$ that could be conveyed to the destination by a random switch at the relay.
If we neglect the term $\log(2)$ we can achieve the upper bound in~\eqref{eq:cs lda} with the scheme shown in Figs.~\ref{fig:HDRCachlindetch relayreceivers} and~\ref{fig:HDRCachlindetch relaysends} for the case $\min\{\beta_{\rm sr},\beta_{\rm rd}\}>\beta_{\rm sd}$, which is the case where the upper bound differs from direct transmission, i.e., $X_r=0$. 
In Phase~I/Fig.~\ref{fig:HDRCachlindetch relayreceivers} the relay listens and the source sends $b_1$ (of length $\beta_{\rm sd}$ bits) directly to the destination and $b_2$ (of length $\beta_{\rm sr}-\beta_{\rm sd}$ bits) to the relay; note that $b_2$ is below the noise floor at the destination; the duration of Phase~I is $\gamma$, hence the relay has accumulated $\gamma(\beta_{\rm sr}-\beta_{\rm sd})$ bits to forward to the destination.
In Phase~II/Fig.~\ref{fig:HDRCachlindetch relaysends} the relay forwards the bits learnt in Phase~I to the destination by `repackaging' them into $a$ (of length $\beta_{\rm rd}-\beta_{\rm sd}$ bits); the source keeps sending a new $b_1$ (of length $\beta_{\rm sd}$ bits) directly to the destination; note that $a$ does not interfere at the destination with $b_2$; the duration of Phase~II is such that all the bits accumulated in Phase~I can be delivered to the destination, that is
\begin{align*}
\gamma(\beta_{\rm sr}-\beta_{\rm sd}) = (1-\gamma)(\beta_{\rm rd}-\beta_{\rm sd}),
\end{align*}
which gives precisely the optimal $\gamma^{*}_{\rm LDA}$. The total number of bits decoded at the destination is 
\begin{align*}
1 \cdot \beta_{\rm sd} + \gamma^{*}_{\rm LDA} \cdot (\beta_{\rm sr}-\beta_{\rm sd}),
\end{align*}
which gives precisely the optimal gDoF for the half-duplex channel in~\eqref{eq:dofHDrelay}.

\begin{remark}
The HD optimal strategy in Figs.~\ref{fig:HDRCachlindetch relayreceivers} and~\ref{fig:HDRCachlindetch relaysends} should be compared with the FD optimal strategy in Fig.~\ref{fig:FDRCachlindetch}. In Fig.~\ref{fig:FDRCachlindetch}, in a given time slot $t$, the source sends $b_1[t]$ (of length $\beta_{\rm sd}$ bits) directly to the destination and $b_2[t+1]$ (of length at most $\beta_{\rm sr}-\beta_{\rm sd}$ bits) to the relay; the relay decodes both $b_1[t]$  and $b_2[t+1]$ and forwards $b_2[t+1]$ in the next slot; in slot $t$ the relay sends $b_2[t]$ (of length at most $\beta_{\rm rd}-\beta_{\rm sd}$ bits) to the destination; the number of bits the relay forwards must be the minimum among the number of bits the relay can decode (given by $\beta_{\rm sr}-\beta_{\rm sd}$) and the number of bits that can be decoded at the destination without harming the direct transmission from the source (given by $\beta_{\rm rd}-\beta_{\rm sd}$). Therefore, the total number of bits decoded at the destination is
\[
\beta_{\rm sd} + \min\{\beta_{\rm rd}-\beta_{\rm sd},\beta_{\rm sr}-\beta_{\rm sd}\},
\]
which gives precisely the optimal gDoF for the full-duplex channel in~\eqref{eq:dofFDrelay}
\end{remark}

\begin{remark}
Fig.~\ref{fig:ldahdfig} compares the capacities of the FD and HD LDA channels; it also shows some achievable rates for the HD LDA channel. In particular, 
the capacity of the FD channel is given by~\eqref{eq:dofFDrelay} (dotted black curve labeled ``FD''),
the capacity of the HD channel is given by~\eqref{eq:exact capacity of LDA HD RC} (solid black curve labeled ``HD'' obtained with the optimal $p_0^*$ in Appendix~\ref{app:exact capacity of LDA HD RC}) and its upper bound by~\eqref{eq:cs lda} (red curve labeled ``HDlda upper'').
For comparison we also show the performance when the source uses i.i.d. Bernoulli$(1/2)$ bits and the relay uses one of the following strategies:
i.i.d. Bernoulli$(q)$ bits and random switch (blue curve labeled ``HDiid q+rand'' obtained by numerically optimizing $q\in[0,1]$),
i.i.d. Bernoulli$(1/2)$ bits and random switch (green curve labeled ``HDiid 1/2+rand'' obtained with $p_0=1/L$ in Appendix~\ref{app:exact capacity of LDA HD RC}), and
i.i.d. Bernoulli$(1/2)$ bits and deterministic switch (magenta curve labeled ``HDiid 1/2+det'' and given by $\beta_{\rm sd}+\min\{\gamma[\beta_{\rm sr}-\beta_{\rm sd}]^+,(1-\gamma)[\beta_{\rm rd}-\beta_{\rm sd}]^+\}$).
We can draw conclusions from Fig.~\ref{fig:ldahdfig}:
\begin{itemize}
\item
With deterministic switch: i.i.d. Bernoulli$(1/2)$ bits for the relay are optimal but this choice is quite far from capacity (magenta curve vs. solid black curve); this choice however is at most one bit from optimal (magenta curve vs. red curve).
\item
With random switch: the optimal input distribution for the relay is not i.i.d. bits; i.i.d. inputs incurs a rate loss (blue curve vs. solid black curve); if in addition we insist on i.i.d. Bernoulli$(1/2)$ bits for the relay we incur a further loss (green curve vs. blue curve). 
\end{itemize}
This shows that for optimal performance the relay inputs are correlated and that random switch should be used.
\end{remark}

\subsection{An achievable strategy inspired by the LDA}
\label{sec:ach simple}

We can mimic the LDA strategy in Section~\ref{sec:achievLDA} for the G-HD-RC as follows. 
We assume $S < C$,  otherwise we use direct transmission to achieve $R=\log(1+S)$.
The transmission is divided into two phases:
\begin{itemize}

\item
Phase~I of duration $\gamma$: the transmit signals are
\begin{align*}
   X_s[1]   &= \sqrt{1-\delta}X_{b_1[1]} + \sqrt{\delta} X_{b_2}, \ \delta = \frac{1}{1+S}, 
\\ X_r[1] &= 0. 
\end{align*}

The relay applies successive decoding of $X_{b_1[1]}$ followed by $X_{b_2}$ from 
\begin{align*}
   Y_r[1] &= \sqrt{C} \ \sqrt{1-\delta}X_{b_1[1]}+ \sqrt{C} \ \sqrt{\delta} X_{b_2} + Z_r[1],
\end{align*}
which is possible if (rates are normalized by the total duration of the two phases) 
\begin{align}
  R_{b_1[1]}&\leq \gamma \log\left(1+C \right)-\gamma \log\left(1+C \frac{1}{1+S}\right)
  \nonumber
\\R_{b_2}   &\leq \gamma \log\left(1+C \frac{1}{1+S}\right).
  \label{r b2  phase 1}
\end{align}

The destination decodes $X_{b_1[1]}$ treating $X_{b_2}$ as noise from 
\begin{align*}
 Y_d[1] &= \sqrt{S} \ \sqrt{1-\delta}X_{b_1[1]} + \sqrt{S} \ \sqrt{\delta} X_{b_2} + Z_d[1],
\end{align*}
which is possible if 
\begin{align}
  R_{b_1[1]} &\leq \gamma \log\left(1+S\right)- \gamma \log\left(1+S \frac{1}{1+S}\right).
  \label{r b1 t phase 1}
\end{align}

Finally, since we assume $S < C$, Phase~I is successful if~\eqref{r b2  phase 1} and~\eqref{r b1 t phase 1} are satisfied.

\item
Phase~II of duration $1-\gamma$: the transmit signals are
\begin{align*}
   X_s[2] &= X_{b_1[2]}
\\ X_r[2] &= X_{b_2} 
\end{align*}
The destination applies successive decoding of $X_{b_2}$ (by exploiting also the information about $b_2$ that it gathered in the first phase) followed by $X_{b_1[2]}$ from 
\begin{align*}
 Y_d[2] &= \sqrt{S} X_{b_1[2]} + \eu^{+\jj \theta} \sqrt{I}  X_{b_2} + Z_d[2],
\end{align*}
which is possible if 
\begin{align}
  &R_{b_2}\! \!\leq (1\!-\!\gamma) \log\left(1\!+\!\frac{I}{1\!+\!S}\right)\!+\! \gamma \log \left( 1\!+\!\frac{S}{1\!+\!S} \right)
  \label{r b2  phase 2}
\\&R_{b_1[2]} \leq (1-\gamma) \log(1+S).
  \label{r b1 t phase 2}
\end{align}

\item
By imposing that the rate $R_{b_2}$ is the same in both phases, that is, that~\eqref{r b2  phase 1} and~\eqref{r b2  phase 2} are equal, we get that $\gamma$ should due chosen equal to $\gamma^*$
\begin{align*}
\gamma^* 
  &\!=\! \frac{ \log\left(1+\frac{I}{1+S}\right)}{ \log\left(1+\frac{I}{1+S}\right)\!+\!\log\left(1+\frac{C}{1+S}\right)\!-\!\log \left( 1+\frac{S}{1+S} \right)}.
\end{align*}
Note that $\gamma^*  \to \gamma^{*}_{\rm LDA}$ as SNR increases. Moreover we give here an explicit closed form expression for the optimal duration of the time the relay listens to the channel.

The rate sent directly from the source to the destination, that is, the sum of~\eqref{r b1 t phase 1} and~\eqref{r b1 t phase 2}, is
\begin{align*}
R_{b_1[1]} + R_{b_1[2]} 
=\log(1+S) - \underbrace{\gamma^* \log\left( 1+ \frac{S}{1+S} \right)}_{\in[0,\log(2)]}.
\end{align*}

Therefore the total rate decoded at the destination through the two phases is
\begin{align*}
&R_{b_1[1]} + R_{b_1[2]} + R_{b_2}
= r^{\rm(LDA-HD)} \ \text{in~\eqref{eq:lda inner bound}}.
\end{align*}

We notice that the rate expression for $r^{\rm(LDA-HD)}$ in~\eqref{eq:lda inner bound}, which was derived under the assumption $C>S$, is valid for all $C$ since for $C<S$ it reduces to direct transmission from the source to the destination.

\end{itemize}

\section{Single Relay Networks: Analytical Gaps}
\label{sec:Analytical Gaps}
In the previous sections we described upper and lower bounds to determine the gDoF of the G-HD-RC.
Here we show that the same upper and lower bounds are to within a constant gap of one another thereby concluding the proof of Theorem~\ref{thm:main 1 relay}.
We consider both the case of random switch and of deterministic switch for the relay.

\begin{proposition}\label{prop:gap pdf rand}
[PDF and random switch]
PDF with random switch is optimal to within \gappdf~bit.
\end{proposition}
\begin{IEEEproof}
The proof can be found in Appendix~\ref{app:prop:gap pdf rand}.
\end{IEEEproof}

\begin{proposition}\label{prop:gap pdf det}
[PDF and deterministic switch]
PDF with deterministic switch is optimal to within \gappdf~bit.
\end{proposition}
\begin{IEEEproof}
The proof can be found in Appendix~\ref{app:prop:gap pdf det}.
\end{IEEEproof}

\begin{proposition}\label{prop:gap lda det}
[LDA (deterministic switch)]
LDA is optimal to within \gaplda~bits.
\end{proposition}
\begin{IEEEproof}
The proof can be found in Appendix~\ref{app:prop:gap lad det}.
\end{IEEEproof}

\bigskip
We conclude this section with a discussion on the gap that can be obtained with NNC. 
The NNC strategy is a network generalization of the CF. It has been proposed for general memoryless networks and it is optimal to within a constant gap for full-duplex multicast networks with an arbitrary number of relays, where the gap grows linearly with the number of relays~\cite{nncLim}. In the case with only one relay, NNC reduces to the classical CF~\cite[Remark 18.6]{book:ElGamalKim2012} and represents a good alternative to the PDF especially in the case when the link between the source and the relay is weaker than the direct link.
The NNC rate is presented in Appendix~\ref{app: inner NNC bounds}.
By using Remark~\ref{rem:simple nnc} in Appendix~\ref{app: inner NNC bounds} we have

\begin{proposition}\label{prop:gap nnc det}
[NNC and deterministic switch]
NNC with deterministic switch is optimal to within \gapnnc~bits.
\end{proposition}
\begin{IEEEproof}
The proof can be found in Appendix~\ref{app:prop:gap nnc det}.
\end{IEEEproof}

%

\section{Single Relay Networks: Numerical Gaps}
\label{sec:Numerical Gaps}
In this section we show that the gap results obtained in Section~\ref{sec:Analytical Gaps} are pessimistic and are due to crude bounding in both the upper and lower bounds, which was necessary in order to obtain rate expressions that could be handled analytically. In order to illustrate our point, we first consider a relay network without the source-destination link, that is, with $S=0$, in Section~\ref{subsec:1relay S=0} and then we show that the same observations are valid for any network in Section~\ref{subsec:1relay anyS}.

\subsection{Single Relay Networks without a Source-Destination Link, a.k.a. Diamond Networks with One Relay}
\label{subsec:1relay S=0}

\paragraph{Upper Bound}
We start by showing that the (upper bound on the) cut-set upper bound in~\eqref{eq:cuset for gap pdf fixed and numerical} can be improved upon. Note that we were not able to evaluate the actual cut-set upper bound in~\eqref{eq:cuset for gap pdf random} so we further bounded it as in~\eqref{eq:cuset for gap pdf fixed and numerical}, which for $S=0$ reduces to
\begin{align*}
r^{\rm(CS-HD)}|_{S=0} 
  =\max_{\gamma\in[0,1]}
\min & \left\{
\mathcal{H}(\gamma)\! +\! (1\!-\!\gamma)\log\left(1\!+\!\frac{I}{1\!-\!\gamma}\right), \right. \nonumber
\\& \left. \gamma\log\left(1+\frac{C}{\gamma}\right)
\right\}.
\end{align*}
The capacity of the G-FD-RC for $S=0$ is known exactly and is given by the cut-set upper bound
\begin{align*}
C^{\rm(FD)}|_{S=0} = \log\left(1+\min\{C,I\}\right).
\end{align*}
$C^{\rm(FD)}$ is a trivial upper bound for the capacity of the G-HD-RC.
Now we show that our upper bound $r^{\rm(CS-HD)}|_{S=0}$ can be larger than $C^{\rm(FD)}|_{S=0}$.
For the case $C = 15/2 > I = 3/2$ 
we have
\begin{align*}
r^{\rm(CS-HD)}|_{S=0}
  &\!\geq \! \min \! \left\{
\mathcal{H}\!\left(\!\frac{1}{2}\right) \!+\! \frac{1}{2}\log\left(1\!+\!2 I\right), \frac{1}{2}\!\log\left(1\!+\!2 C \!\right)
\right\}
\\&= \log(4) > C^{\rm(FD)}|_{S=0} = \log\left(2.5\right).
\end{align*}
The reason why the capacity of the FD channel can be smaller than our upper bound $r^{\rm(CS-HD)}|_{S=0}$ is the crude bound $I(S_r;Y_d) \leq H(S_r) = \mathcal{H}(\gamma)$. As mentioned earlier, we needed this bound in order to have an analytical expression for the upper bound.
Actually for $S=0$ the cut-set upper bound in~\eqref{eq:cuset for gap pdf random} is tight, as we show next.

\paragraph{Exact capacity with PDF}
\begin{theorem}
In absence of direct link between the source and the destination PDF with random switch achieves the cut-set upper bound.
\end{theorem}
\begin{IEEEproof}
With $S=0$, the cut-set upper bound in~\eqref{eq:cuset for gap pdf random} and the PDF lower bound in~\eqref{eq:inner for gap pdf random} are the same (see also Appendix~\ref{app:prop:gap pdf rand} with $S=0$). \end{IEEEproof}

\paragraph{Improved gap for the LDA Lower Bound}
Despite knowing the capacity expression for $S=0$, its actual evaluation is elusive as it is not clear what the optimal input distribution $P^*_{X_s,X_r,S_r}$ in~\eqref{eq:cuset for gap pdf random} is. For this reason we next specialized the LDA strategy to the case $S=0$ and evaluate its gap from the (upper bound on the) cut-set upper bound in~\eqref{eq:cuset for gap pdf fixed and numerical}.

The LDA achievable rate in~\eqref{eq:lda inner bound} with $S=0$ is given by
\begin{align*}
r^{\rm(LDA-HD)}|_{S=0} = \max_{\gamma\in[0,1]} \min \{& \gamma \log \left( 1+C \right)
\\& (1-\gamma) \log \left( 1+I \right)  \}.
\end{align*}
and its gap from the outer bound can be reduced from \gaplda~bits to about 1.5~bits since
\begin{align*}
\mathsf{GAP} &\leq r^{\rm(CS-HD)}|_{S=0} - r^{\rm(LDA-HD)}|_{S=0}
\\& \leq \max_{\gamma\in[0,1]} \left \{\gamma\log\left(1+\frac{C}{\gamma}\right)-\gamma \log \left( 1+C \right), \right.
\\& \left. \quad \mathcal{H}(\gamma) \!+\!(1-\gamma)\log\left(1\!+\!\frac{I}{1\!-\!\gamma}\right)\!-\!(1\!-\!\gamma) \log \left( 1\!+\!I \right)\right \}
\\& \leq \max_{\gamma\in[0,1]} \left \{ \gamma\log\left( \frac{1}{\gamma} \right), \mathcal{H}(\gamma) + (1-\gamma) \log \left( \frac{1}{1-\gamma} \right) \right \}
\\& = \!\max_{\gamma\in[0,1]} \!\left\{
\mathcal{H}(\gamma) \!+\! (1-\gamma) \log \left( \frac{1}{1\!-\!\gamma} \right)
\right\} \!=\! 1.5112 \ \rm{bits}. 
\end{align*}
Note that the actual gap is even less than 1.5~bits. By numerically evaluating the difference between $\min\{C^{\rm(FD)},r^{\rm(CS-HD)}\}|_{S=0}$ and $r^{\rm(LDA-HD)}|_{S=0}$ we found that the gap is at most 1.11~bits. 

\paragraph{Numerical gaps with deterministic switch}
Similarly, by numerical evaluations one can find that the PDF strategy with deterministic switch in Remark~\ref{rem:simple pdf}-Appendix~\ref{app: proof of prop:inner upper bounds} and the NNC strategy with deterministic switch in Remark~\ref{rem:simple nnc}-Appendix~\ref{app: inner NNC bounds} are to within 0.80~bits and 1.01~bits, respectively, of the improved upper bound. Notice that in these cases there is no information conveyed by the relay to the destination through the switch. Further reductions in the gap with random switch are discussed next for a general network.

Fig.~\ref{fig:diamond1relayhdfig} shows different upper an lower bounds for the G-HD-RC for $S=0, \ C=15, \ I=3$ vs $\gamma=\mathbb{P}[S_r=0]$. We see that the cut-set upper bound exceeds the capacity of the G-FD-RC (maximum of the solid black curve vs. dashed black curve). Different achievable strategies are also shown, whose order from the most performing to the least performing is:
PDF with random switch (red curve, 1.913 bits/ch.use),
PDF with deterministic switch (blue curve, 1.702 bits/ch.use), 
NNC with random switch (cyan curve, 1.446 bits/ch.use),  
NNC with deterministic switch (magenta curve, 1.402 bits/ch.use), and 
LDA (green curve, 1.333 bits/ch.use).  
In this particular setting, the maximum rate using the NNC strategy with random switch (cyan curve, 1.446 bits/ch.use) is achieved for  
$\mathbb{P}[Q=0,S_r=0]=0,\mathbb{P}[Q=0,S_r=1]=0.33,\mathbb{P}[Q=1,S_r=0]=0.45,\mathbb{P}[Q=1,S_r=1]=0.22$. This is due to the absence of the direct link ($S=0$) between the source and the destination. Actually, since the source can communicate with the destination only through the relay, it is necessary a coordination between the transmissions of the source and those of the relay. This coordination is possible thanks to the time-sharing random variable $Q$, i.e. when $Q=0$ the source stays silent, while when $Q=1$ the source transmits.

\subsection{Single Relay Network with a Source-Destination Link}
\label{subsec:1relay anyS}
Although the considerations in Section~\ref{subsec:1relay S=0} were for a relay channel without a source-destination link, they are valid in general.

Fig.~\ref{fig:PDFbetter} and Fig.~\ref{fig:NNCbetter} show the rates achieved by using the different achievable schemes presented in the previous sections with $S>0$. In Fig.~\ref{fig:PDFbetter} the channel conditions are such that the PDF strategy outperforms the NNC, while in Fig.~\ref{fig:NNCbetter} the opposite holds. In Fig.~\ref{fig:PDFbetter} the PDF strategy with random switch (red curve, 11.66 bits/ch.use) outperforms both the NNC with random switch (cyan curve, 11.11 bits/ch.use) and the PDF with deterministic switch (blue curve, 11.4 bits/ch.use); then the PDF with deterministic switch outperforms the NNC with deterministic switch (magenta curve, 10.94 bits/ch.use), which is also encompassed by the NNC with random switch.
Differently from the case without direct link, we observe that the maximum NNC rates both in Fig.~\ref{fig:PDFbetter} and in Fig.~\ref{fig:NNCbetter} are achieved with the choice $Q=\emptyset$, i.e. the time-sharing random variable $Q$ is a constant. This is due to the fact that the source is always heard by the destination even when the relay transmits so there is no need for the source to remain silent when the relay sends.

In Fig.~\ref{fig:actuallgapI0zero} we consider the case of deterministic switch.
Fig.~\ref{fig:actuallgapI0zero} shows, as a function of SNR for $\beta_{\rm sd} =1, (\beta_{\rm rd},\beta_{\rm sr})\in[0,2.4]$, the maximum gap between the cut-set upper bound $r^{\rm(CS-HD)}$ in~\eqref{eq:cuset for gap pdf fixed and numerical} and the following lower bounds with deterministic switch: the PDF lower bound obtained from $r^{\rm(PDF-HD)}$ in~\eqref{eq:inner for gap pdf fixed and numerical} with $I_0^{\rm(PDF)}=0$, the NNC lower bound in Remark~\ref{rem:simple nnc} in Appendix~\ref{app: inner NNC bounds}, and the LDA lower bound in~\eqref{eq:lda inner bound}.
From Fig.~\ref{fig:actuallgapI0zero} we observe that the maximum gap with PDF is of 1~bit as in Proposition~\ref{prop:gap pdf det}, but with NNC is around 1.52~bits and with LDA is around 1.59~bits, which are lower than the analytical  gaps we found in Propositions~\ref{prop:gap nnc det} and~\ref{prop:gap lda det}, respectively.

The lower bounds can be improved upon by considering that information can be transmitted through a random switch for the relay.
However, this improvement depends on the channel gains. If the information cannot be routed through the relay because $\min\{C,I\}\leq S$, then the system cannot exploit the randomness of the switch, and so $I_0^{\rm PDF} = 0$ and $I_0^{\rm NNC} = 0$ are approximately optimal (in this case the relay can remain silent). 
For this reason the maximum numerical gap obtained with a random switch coincides with the one obtained with a deterministic switch, as there are channel conditions for which random switch is not necessary.
This behavior for the PDF strategy is represented in Fig.~\ref{fig:actuallgapPDF}. In this figure we numerically evaluate the difference between the analytical gap, i.e., the one computed with $I_0^{\rm PDF} = 0$, and the numerical one, i.e., computed with $I_0^{\rm{opt}}$ (actual value of $I_0^{\rm{PDF}}$), at a fix $\rm{SNR}=20\rm{dB}$ and by varying $(\beta_{\rm rd},\beta_{\rm sr})$. We observe that when the information cannot be conveyed through the relay, i.e., $\min \left \{ \beta_{\rm rd},\beta_{\rm sr} \right \}\leq 1$, then $I_0^{\rm PDF}=0$ is optimal, since the information only flows through the direct link.     
In Fig.~\ref{fig:gapPDFdetrandom} the channel channel gains are set such that the use of the relay increases the gDoF of the channel ($\beta_{\rm sd} =1$ and $(\beta_{\rm rd},\beta_{\rm sr})\in[1,2.4]$). Here the relay uses PDF. We observe that we have a further improvement in terms of gap by using a random switch (blue curve) instead of using a deterministic switch (red curve). We notice that at high SNR, where the gap is maximum, this improvement is around 0.1~bits. As mentioned earlier, the rate advantage of random switch over deterministic switch depends on the channel gains.


\section{Networks with Multiple Relays}
\label{sec:multirelay}
In this section we extend our gDoF and gap results to general HD-MRC. Similarly to the the full-duplex case~\cite{nncLim}, our main result is that NNC is optimal to within a constant gap for the HD-MRC where the gap is a function of the number of relays.
 
\subsection{Network Model}
In this model we have $K \geq 3$ nodes, i.e., one transmitter/node~$1$, one receiver/node~$K$ and $K-2$ relays with indices $2,\ldots,K-1$. Each node is equipped with a single antenna and is subject to an average power constraint, which we set to $1$ without loss of generality, i.e. $\mathbb{E} \left [ |X_{k}|^2   \right ] \leq 1 \in \mathbb{R}^+$, $k \in [1:K-1]$. The system is described by the input/output relationship
\begin{subequations}
\begin{align}
&\mathbf{Y} 
=
(\mathbf{I}-\mathbf{S}) 
\mathbf{H} 
\mathbf{S} 
\
\mathbf{X} 
+
\mathbf{Z}
\\&\mathbf{Z}=[Z_{1},\ldots,Z_{K}]^T \sim \mathcal{N}(\mathbf{0},\mathbf{I})
\\&\mathbf{Y}=[Y_{1},\ldots,Y_{K}]^T  \in \mathbb{C}^{K}
\\&\mathbf{X}=[X_{1},\ldots,X_{K}]^T\in \mathbb{C}^{K} : \nonumber
\\& \quad \quad \mathbb{E}[|X_k|^2] \leq 1 \ {\rm for}\ k\in[1:K-1],
\\&\mathbf{H} \in \mathbb{C}^{K\times K}
\\& \mathbf{S} = {\rm diag}\{S_1,\ldots,S_{K}\} : \nonumber
\\& \quad S_1 = 1, \ S_k\in\{0,1\} \ {\rm for}\ k\in[2:K-1], \ S_K=0,
\end{align}
\label{eq:channMulti}
\end{subequations}
where the vector $\mathbf{S}$ represents the state of the nodes, either receive ($S = 0$) or transmit ($S = 1$). The channel matrix $\mathbf{H} \in \mathbb{C}^{K\times K}$ is constant and therefore known to all terminals. The entry $h_{ij}$ with $(i,j) \in [1:K]^2$ represents the channel between source $j$ and destination $i$.
Without loss of generality we assume that the noises $\mathbf{Z}$ are zero-mean jointly Gaussian and with unit power. Furthermore we assume that the noises are iid $\mathcal{N}(0,1)$, this is however not without loss of generality~\cite{ZhangCorrNoise}.

The capacity of the channel described in \eqref{eq:channMulti} is not known in general. Here we show that a scheme based on the NNC strategy achieves the capacity within a constant gap for any number of relays and for any choice of channel parameters. As for the FD case, the gap is found to be a function of the number of relays. We propose a different and simpler bounding technique than that of~\cite{nncLim}, which might overestimate the actual gap between inner and outer bound.

\subsection{Capacity to within a constant gap}
Our main result is
\begin{theorem}
\label{thm:many relays}
The cut-set upper bound for the half-duplex multi-relay network is achievable to within
\begin{align}
\label{eq:GapMultipleRelay}
\mathsf{GAP} \leq \max_{\ell\in[0:K-2]}\{&\min\{ 1+\ell, K-1-\ell\}  \log\left(1+\ell\right) \nonumber
\\& +\min \{1+3\ell,\ell+K-1 \}\}
\end{align}
bits per channel use.
\end{theorem}

\begin{IEEEproof}
The proof can be found in Appendix~\ref{app:prop:gap multi relay}.
\end{IEEEproof}

For high value of $K$, i.e. $K>>1$, the optimal value of $\ell$ in \eqref{eq:GapMultipleRelay} is $\ell \cong \frac{K-2}{2}$ and the corresponding limit gap becomes 
\begin{align}
\label{eq:GapMultipleRelayAsympt}
\mathsf{GAP} \cong \frac{K}{2} \log(4K).
\end{align}
Fig.~\ref{fig:GapVaryingUsers} shows the trend of the gap in \eqref{eq:GapMultipleRelay} as a function of the number of nodes $K$ (blue curve). The red curve represents the limit behaviour of the gap in \eqref{eq:GapMultipleRelay}.

Note that by applying the above gap-result to the case $K=3$  we obtain $\mathsf{GAP}\leq 4~{\rm bits}$
which is a much larger gap than \gapnnc~bits we found for the case of one relay.

A smaller gap may be obtained by computing tighter bounds. This can be accomplished by several means. For example:
\begin{itemize}
\item
By using more complex and sophisticated bounding techniques: in~\cite{nncLim} an upper bound on the water-filling power allocation for a general MIMO channel was derived. This more involved upper bound could be used here to obtain a smaller gap. We note that our bounding technique applied to the FD-MRC gives a gap of order $ \frac{K}{2}\log(2K)$, which is larger that $0.63 K$ found in~\cite{nncLim}.

\item
By using an achievable strategy based on PDF, which in a single relay case gives a smaller gap than NNC.  However, PDF seems not to be easily extended to networks with an arbitrary number of relays, which is the main motivation we consider NNC here. 

\item
By deriving tighter bounds on specific network topologies: in~\cite{arXiv:1207.5660} it is found that for a FD diamond network with $K$ relays the gap is of the order $\log(K)$, rather than linear in $K$~\cite{nncLim}. Moreover, for a symmetric FD diamond network with $K$ relays the gap does not depend on the number of relays and is upper bounded by 2~bits. The key difference between a general multi-relay network and a diamond network is that for each subset $\mathcal{A}$ we have ${\rm Rank}[\mathbf{H}_{\mathcal{A},s}]=2$, i.e., the rank of a generic channel sub-matrix does no longer depend on the cardinality of $\mathcal{A}$.  Based on this observation we have
\begin{proposition}\label{prop:hd diamond}
The cut-set upper bound for the gaussian half-duplex diamond network with $\left(K-2\right)$ relays is achievable to within 
\begin{align*}
&\mathsf{GAP} \leq (K-2)\log(2)+4\log(K)+2\log({\rm e}/2).
\end{align*}
Moreover, if the conjecture in~\cite{Fragouli2012} holds then the gap above could be decreased to
\begin{align*}
&\mathsf{GAP} \leq 5\log(K)+2\log({\rm e}/2).
\end{align*}
\end{proposition}
\begin{IEEEproof}
The proof can be found in Appendix~\ref{app:hd diamond}.
\end{IEEEproof}
\end{itemize}

\subsection{Example: Fully connected network with $K=4$}
To gain insights into how relays are best utilized, 
we consider a network with two relays, i.e. $K=4$ nodes. 
In particular we highlight under which channel conditions the gDoF performance is enhanced by exploiting both relays rather than using only the best one. We also compare the loss that incurs by using HD with respect to FD.
Let parameterize the channel gains as
\begin{align*}
\left[\frac{\log(|h_{ij}|^2)}{\log(\mathsf{SNR})}\right]_{(i,j)\in[1:4]^2}
=
\begin{pmatrix}
* & * & * & * &  \\
\alpha_{s1} & * & \beta_1 & * \\ 
\alpha_{s2} & \beta_2 & * & * \\ 
1 & \alpha_{1d} & \alpha_{2d} & * \\ 
\end{pmatrix}
\end{align*}
where $*$ denotes entries that do not matter; this is so because the source node never listens to the channel (first row), the destination node never transmits (last column), and a relay can remove the `self-interference' (main diagonal). Note that the direct link from the source to the destination has gain $\mathsf{SNR}$ and all other channel gains are expressed with reference to it.

\paragraph{Full Duplex}
In the following the set $\mathcal{A}$ indicates the relays that lie on the source/node~1 side of the cut. The cut-set upper bound for the FD-MRC~\cite{nncLim} gives
\begin{align*}
  &\mathcal{A}=\emptyset: && I_1 : = I(X_1       ; Y_4,Y_2,Y_3| X_2,X_3) 
\\&                       && \leq \log(1+|h_{41}|^2+|h_{21}|^2+|h_{31}|^2) 
\\&\mathcal{A}=\{2\}    : && I_2:= I(X_1,X_2    ; Y_4,    Y_3|     X_3) 
\\&                       && \leq \log|\mathbf{I}_2 +\mathbf{H}_1 \mathbf{H}_1^H|+\log(2)
\\&                       && = \log \left( 1+A \right)+2\log(2)
\\&\mathcal{A}=\{3\}    : && I_3: =I(X_1,    X_3; Y_4,Y_2    | X_2    ) 
\\&                       && \leq \log|\mathbf{I}_2 +\mathbf{H}_2 \mathbf{H}_2^H|+\log(2)
\\&                       && = \log \left( 1+B \right)+2\log(2)
\\&\mathcal{A}=\{2,3\}  : && I_4 := I(X_1,X_2,X_3; Y_4       |\emptyset) 
\\&                       && \leq \log(1+(|h_{41}|+|h_{42}|+|h_{43}|)^2),
\end{align*}
where 
\begin{align*}
&A=|h_{41}|^2+|h_{42}|^2+|h_{31}|^2+|h_{32}|^2+ |h_{31}h_{42}-h_{32}h_{41}|^2, 
\\&B=|h_{41}|^2+|h_{43}|^2+|h_{21}|^2+|h_{23}|^2+ |h_{21}h_{43}-h_{23}h_{41}|^2,
\\&\mathbf{I}_2 =
\begin{bmatrix}
1 & 0  \\
0 & 1  \\ 
\end{bmatrix}
, \
\mathbf{H}_1 = 
\begin{bmatrix}
h_{31} & h_{32}  \\
h_{41} & h_{42}  \\ 
\end{bmatrix}
\quad \text{and} \
\mathbf{H}_2 = 
\begin{bmatrix}
h_{21} & h_{23} \\
h_{41} & h_{43}
\end{bmatrix}.
\end{align*}
Thus for the full-duplex case, the cut-set bound is given by
\begin{align*}
r^{\rm(CS-FD)} := \min \left \{I_1,I_2,I_3,I_4 \right \}
\end{align*}
which at high $\mathsf{SNR}$ gives the following gDoF (achievable to within a constant gap~\cite{nncLim})
\begin{align*}
\mathsf{d}_{K=4}^{\rm(FD)} 
  &=\lim_{\mathsf{SNR}\to+\infty} \frac{r^{\rm(CS-FD)}}{\log(1+\mathsf{SNR})}
\\&= \min \Big\{
\max \left\{ 1,\alpha_{s1},\alpha_{s2} \right \},
\max \left\{ \alpha_{s2}+\alpha_{1d},\beta_2+1 \right \},
\\&\quad \quad \quad
\max \left\{ \alpha_{s1}+\alpha_{2d},\beta_1+1 \right \},
\max \left\{ 1,\alpha_{1d},\alpha_{2d} \right \} 
\Big \}.
\end{align*}

In the following we are interested in identifying cases for which 
\begin{align*}
\mathsf{d}_{K=4}^{\rm(FD)} >
\mathsf{d}_{K=4,\text{best relay}}^{\rm(FD)}:=&
\max\big\{ 
1, \min\{\alpha_{s1},\alpha_{1d}\}, 
\\& \quad \quad \quad \min\{\alpha_{s2},\alpha_{2d}\}
\big\},
\end{align*}
where $\mathsf{d}_{K=4,\text{best relay}}^{\rm(FD)}$ is the gDoF that can be obtained by selecting the relay which achieves the highest gDoF in~\eqref{eq:dofFDrelay} while leaving the other silent.
From the expression for $\mathsf{d}_{K=4}^{\rm(FD)}$ we immediately see that in order to have $\mathsf{d}_{K=4}^{\rm(FD)} > 1$, i.e., better than direct transmission with the two relays silent,  we need
\begin{align*}
\max \left\{\alpha_{s1},\alpha_{s2} \right \} > 1, \\
\max \left\{\alpha_{s2}+\alpha_{1d}-1,\beta_2 \right \}>0, \\ 
\max \left\{\alpha_{s1}+\alpha_{2d}-1,\beta_1 \right \}>0, \\
\max \left\{\alpha_{1d},\alpha_{2d} \right \} > 1.
\end{align*}
We therefore distinguish the following cases:
\begin{enumerate}

\item \label{item k=4 equal k=3: case 1}
if $\alpha_{s1}> \max\{1,\alpha_{s2}\}$ and $\alpha_{1d}> \max\{1,\alpha_{2d}\}$ then
\begin{align*}
\mathsf{d}_{K=4}^{\rm(FD)}= \min\{\alpha_{s1},\alpha_{1d}\} = \mathsf{d}_{K=4,\text{best relay}}^{\rm(FD)}  > 1;
\end{align*}
therefore this case is not interesting;

\item 
if $\alpha_{s1}> \max\{1,\alpha_{s2}\}$ and $\alpha_{2d}> \max\{1,\alpha_{1d}\}$ then
\begin{align*}
\mathsf{d}_{K=4}^{\rm(FD)} 
= 
\min \Big\{
\alpha_{s1},
\max \left \{ \alpha_{s2}+\alpha_{1d},\beta_2+1 \right \},
\alpha_{2d} 
\Big \}.
\end{align*}
Next consider the following sub-cases:
\begin{enumerate}

\item \label{item k=4 strictly better than k=3: case 1}
If $\alpha_{1d}\leq 1$ or $\alpha_{s2}\leq 1$, 
then $\mathsf{d}_{K=4,\text{best relay}}^{\rm(FD)} < \mathsf{d}_{K=4}^{\rm(FD)}$, hence in this case using both relays can give an unbounded rate improvement over using the best relay if $ \max \left \{ \alpha_{s2}+\alpha_{1d}-1,\beta_2 \right \} > 0$;

\item
If $\alpha_{1d}> 1$ and $\alpha_{s2}> 1$, 
then $\mathsf{d}_{K=4,\text{best relay}}^{\rm(FD)}= \max\big\{ \min\{\alpha_{s1},\alpha_{1d}\}, \min\{\alpha_{s2},\alpha_{2d}\}\big\} > 1$; we distinguish

\begin{enumerate}

\item \label{item k=4 equal k=3: case 2}
If $\alpha_{s1} \leq \alpha_{1d}$ or $\alpha_{2d} \leq \alpha_{s2}$ then $\mathsf{d}_{K=4,\text{best relay}}^{\rm(FD)}=\min\{\alpha_{s1},\alpha_{2d}\}=\mathsf{d}_{K=4}^{\rm(FD)} $, therefore this case is not interesting; 
 
\item \label{item k=4 strictly better than k=3: case 2}
If $\alpha_{s1} > \alpha_{1d}$ and $\alpha_{2d} > \alpha_{s2}$ then $\mathsf{d}_{K=4,\text{best relay}}^{\rm(FD)}=\max\{\alpha_{1d},\alpha_{s2}\} < \mathsf{d}_{K=4}^{\rm(FD)}$, therefore we have a strict improvement by using both relays over using only the best relay.

\end{enumerate}

\end{enumerate}

\item 
Considering cases similar to the two above but with the role of the relays swapped complete the list of possible cases.

\end{enumerate}

An example of network satisfying the conditions in item~\ref{item k=4 strictly better than k=3: case 1} or item~\ref{item k=4 strictly better than k=3: case 2} is given in Fig.~\ref{fig:tworelaynet} where the numerical value on a link represents the SNR exponent on the corresponding link.

\paragraph{Half Duplex}
With HD, each term in the min-function defining the cut-set bound is a convex combination of $2^{K-2}$ terms, one for each possible state of the network. Formally, these terms are as for the FD case but where one replaces the SNR exponent of the link from node~$j$ to node~$i$, given by $\alpha_{ij}$, with $(1-S_i) \alpha_{ij} S_j$, where $(S_i,S_j)\in\{0,1\}^2$ indicates the state of the nodes. Because no additional insight is gained from these terms, we do not report them here. At high SNR the outer bound gives the following gDoF
\begin{align*}
\mathsf{d}_{K=4}^{\rm(HD)} 
=  \max
\min\Big\{
  & \lambda_{00} D_1^{(0)}\!+\!\lambda_{01} D_1^{(1)}\!+\!\lambda_{10} D_1^{(2)}\!+\!\lambda_{11} D_1^{(3)}, 
\\& \lambda_{00} D_2^{(0)}\!+\!\lambda_{01} D_2^{(1)}\!+\!\lambda_{10} D_2^{(2)}\!+\!\lambda_{11} D_2^{(3)}, 
\\& \lambda_{00} D_3^{(0)}\!+\!\lambda_{01} D_3^{(1)}\!+\!\lambda_{10} D_3^{(2)}\!+\!\lambda_{11} D_3^{(3)}, 
\\& \lambda_{00} D_4^{(0)}\!+\!\lambda_{01} D_4^{(1)}\!+\!\lambda_{10} D_4^{(2)}\!+\!\lambda_{11} D_4^{(3)}
\Big\},
\end{align*}
where the maximization is over $\lambda_{ij}$, $(i,j) \in \{ 0,1\}^2$ representing the fraction of time node~2 is in state $S_2=i$ and node~3 is in state $S_3=j$, such that 
$\lambda_{00}+\lambda_{01}+\lambda_{10}+\lambda_{11}=1$ and
\begin{align*}
  & D_1^{(0)} := \max \left \{ 1,\alpha_{s1},\alpha_{s2} \right \},
\\& D_1^{(1)} = D_3^{(0)} := \max \left \{ 1,\alpha_{s1} \right \},
\\& D_1^{(2)} = D_2^{(0)} := \max \left \{ 1,\alpha_{s2} \right \},  
\\& D_1^{(3)} = D_2^{(1)} = D_3^{(2)} = D_4^{(0)} := 1,   
\\& D_2^{(2)} :=  \max  \left \{  \alpha_{s2}+\alpha_{1d}, \beta_2 +1 \right \},  
\\&D_2^{(3)} = D_4^{(2)} := \max \left \{ 1,\alpha_{1d} \right \}, 
\\&D_3^{(1)} := \max \left \{ \alpha_{s1}+\alpha_{2d},\beta_1 +1 \right \},
\\&D_3^{(3)} = D_4^{(1)} :=\max \left \{ 1,\alpha_{2d} \right \},
\\&D_4^{(3)} := \max \left \{ 1,\alpha_{1d},\alpha_{2d} \right \}.
\end{align*}
An analytical closed form solution for the optimal $\{\lambda_{ij}\}$ is complex to find for general channel gains.  However, numerically it is a question of solving a linear program, for which efficient numerical routines exist. We remark that this linear program can be thought as the high SNR solution of the iterative algorithm proposed in~\cite{OngMultiRelay}.

Table~\ref{table:nonlin} shows the gDoF corresponding to the four cases listed in the FD case where using both relays strictly improves over exploiting only the best relay. 
We denote with $\mathsf{d}_{K=4,\text{best relay}}^{\rm(HD)}$ the gDoF that is obtained when the two relays work in HD by selecting the relay which achieves the highest gDoF while leaving the other silent, whose closed form solution is given in~\eqref{eq:dofHDrelay}.
From Table \ref{table:nonlin} we notice that in each case also for the HD case we have $\mathsf{d}_{K=4}^{\rm(HD)} > \mathsf{d}_{K=4,\text{best relay}}^{\rm(HD)}$, as for the FD case. Furthermore, as expected, $\mathsf{d}_{K=4}^{\rm(FD)}>\mathsf{d}_{K=4}^{\rm(HD)}$.

\begin{table*}
\centering 
\caption{gDoF when using both relays is better than using only the best one} 
\label{table:nonlin} 
\begin{tabular}{c c c c c} 
\toprule 
\textbf{Channel Parameters} & \textbf{Full-duplex} & \textbf{Full-duplex} & \textbf{Half-duplex} & \textbf{Half-duplex}\\
\textbf{$\left( \alpha_{s1},\alpha_{s2},\alpha_{1d},\alpha_{2d},\beta_1,\beta_2 \right)$} & \textbf{best relay} & \textbf{both relays} & \textbf{best relay} & \textbf{both relays} \\
\midrule
$\left( 2.5,1.4,0.5,1.8,0.6,0.8 \right)$ & $1.4$ & $1.8$ & $1.267$ & $1.4235$ \\ 
$\left( 2.5,0.3,0.7,1.3,0.4,0.8 \right)$ & $1.0$ & $1.3$ & $1.000$ & $1.2182$ \\
$\left( 1.8,1.2,1.3,2.0,0.7,1.2 \right)$ & $1.3$ & $1.8$ & $1.218$ & $1.5808$ \\
$\left( 1.7,1.1,1.2,1.4,0.4,1.5 \right)$ & $1.2$ & $1.4$ & $1.156$ & $1.3604$ \\
\bottomrule
\end{tabular}
\end{table*}

\section{Conclusions}
\label{sec:conclusion}
In this work we considered the HD-RC, a network where the source communicates with the destination through a relay node. The relay node works in half-duplex, in the sense that it transmits and receives in different time slots. This scenario, with the half-duplex assumption, represents a more practically relevant model compared to the full-duplex case. In particular, it is applicable in practical relaying architectures for 4G cellular networks.

We derived in a close form expression the generalized Degrees-of-Freedom of this system and we show that three schemes achieve the capacity to within a constant gap. The first scheme turns out to be simple both in the encoding and decoding phases and it is inspired by the Linear Deterministic Approximation of the Gaussian noise at high SNR; the second and the third schemes are based on the Partial-Decode-and-Forward and Noisy-Network-Coding strategies.

All these schemes consider both deterministic and random switch at the relay. In the first case the switch is known by all the terminals in the network, while in the second case the randomness that lies into the switch can be exploited to send more information and so to achieve higher rates and consequently lower gaps. We show that random switch is optimal and in some cases, as in the diamond network with only one relay, achieves the exact capacity.

Finally we extend our results to a multi-relay channel and we show that a strategy based on the NNC with random switch at each relay, achieves the capacity to within a constant gap. We also prove that this gap may be even decreased in more structured settings as, for example, the diamond network where there are no source-destination and relay-relay links.


\bibliographystyle{IEEEbib}
\bibliography{IccBib}


\appendices 

\section{Proof of Proposition~\ref{prop:cutset upper bounds}}
\label{app: proof of prop:cutset upper bounds}
\begin{IEEEproof}
An outer bound to the capacity of the memoryless RC is given by the cut-set outer bound~\cite[Thm.16.1]{book:ElGamalKim2012} 
that specialized to our G-HD-RC channel gives
\begin{subequations}
\begin{align}
&C^{\rm(HD-RC)} \nonumber
\\& \leq \!\!\!\!\!\max_{P_{X_s,[X_r,S_r]}}\!\! \!\!
\min\Big\{
I(X_s,[X_r,S_r]; Y_d),
I(X_s; Y_r,Y_d| [X_r,S_r])
\Big\}\label{eq:the one leading to eq:cuset for gap pdf random}
\\&= \max_{P_{X_s,X_r,S_r}}
 \min\Big\{
I(S_r; Y_d) + I(X_s,X_r; Y_d| S_r), \nonumber
\\& \quad \quad \quad \quad \quad \quad \quad \quad I(X_s; Y_r,Y_d| X_r,S_r)
\Big\}
\\& \leq  \max_{P_{X_s,X_r,S_r}}
 \min\Big\{
H(S_r) + I(X_s,X_r; Y_d| S_r), \nonumber
\\& \quad \quad \quad \quad \quad \quad \quad \quad I(X_s; Y_r,Y_d| X_r,S_r)
\Big\}\label{eq:bound on I(Sr;Yd)}
\\&\leq  
\max\min\Big\{
\mathcal{H}(\gamma) + 
 \gamma I_{1} +(1-\gamma)I_{2}, \
 \gamma I_{3} +(1-\gamma)I_{4}
\Big\} \nonumber
\\& =: r^{\rm(CS-HD)}, \label{eq:and finally N-max-h}
\end{align}
\end{subequations}
where the different steps follow since:
\begin{itemize}

\item
We indicate the (unknown) distribution that maximizes~\eqref{eq:the one leading to eq:cuset for gap pdf random} as $P^*_{X_s,X_r,S_r}$ in order to get the bound in~\eqref{eq:cuset for gap pdf random}.

\item
In order to obtain the bound in~\eqref{eq:bound on I(Sr;Yd)} we used the fact that for a discrete binary-valued random variable $S_r$ we have
\begin{align*}
I(S_r; Y_d) = H(S_r)-H(S_r|Y_d) \leq H(S_r) = \mathcal{H}(\gamma)
\end{align*}
for some $\gamma:=\mathbb{P}[S_r=0]\in[0,1]$ that represents the fraction of time the relay listens and where
$\mathcal{H}(\gamma)$ is the binary entropy function in~\eqref{eq:binaty entropy function}.
In~\eqref{eq:and finally N-max-h} the maximization is over the set defined by~\eqref{eq:GHDRC CS gamma}-\eqref{eq:GHDRC CS powers} and is obtained as an application of the `Gaussian maximizes entropy' principle as follows.
Given any input distribution $P_{X_s,X_r,S_r}$, the covariance matrix of $(X_s,X_r)$ conditioned on $S_r$ can be written as
\begin{align*}
{\rm Cov}\left.
\begin{bmatrix}
X_s \\
X_r \\ 
\end{bmatrix}\right|_{S_r=\ell}
=
\begin{bmatrix}
P_{s,\ell} & \alpha_{\ell} \ \sqrt{P_{s,\ell}P_{r,\ell}} \\ 
\alpha_{\ell}^* \ \sqrt{P_{s,\ell}P_{r,\ell}} & P_{r,\ell}\\
\end{bmatrix},
\end{align*}
with $|\alpha_\ell|\leq 1$ for some $(P_{s,0},P_{s,1},P_{r,0},P_{r,1}) \in \mathbb{R}^4_{+}$ satisfying the average power constraint in~\eqref{eq:GHDRC CS powers}. Then, a zero-mean jointly Gaussian input with the above covariance matrix maximizes the different mutual information terms in~\eqref{eq:bound on I(Sr;Yd)}. In particular
\begin{align*}
 & I(X_s,X_r; Y_d|S_r=0) \leq \log \left(1+S P_{s,0}\right) =: I_1,
\\&I(X_s,X_r; Y_d|S_r=1) 
\\& \leq \!  \log \left(1\!+\!S P_{s,1}\!+\!I P_{r,1} \!+\!2|\alpha_1|\sqrt{  S P_{s,1} \ IP_{r,1}}\right) =: I_{2},
\\&I(X_s;Y_r,Y_d|X_r, S_r=0) 
\\& \leq \log\left(1+(C +S)(1-|\alpha_0|^2) P_{s,0}\right) 
 \\&                            \leq \log\left(1+(C +S) P_{s,0}\right)=: I_{3},
\\&I(X_s;Y_r,Y_d| X_r,S_r=1) \leq \log\left(1+S(1-|\alpha_1|^2) P_{s,1}\right)
\\& =: I_{4},
\end{align*}
as defined in~\eqref{eq:def of I1}-\eqref{eq:def of I4} thereby proving the upper bound in~\eqref{eq:cuset for gap pdf fixed and numerical}, which is the same as $r^{\rm(CS-HD)}$ in~\eqref{eq:and finally N-max-h}. This shows the bound in~\eqref{eq:cuset for gap pdf fixed and numerical}.

\item
In order to get to~\eqref{eq:cuset for gap analytical} from~\eqref{eq:cuset for gap pdf fixed and numerical} we let the channel gains be parameterized as in~\eqref{eq:param high snr}. The average power constraints at the source and at the relay given in~\eqref{eq:GHDRC CS powers} can be expressed as follows. Since the source transmits in both phases we define for some $\beta\in[0,1]$
\begin{align*}
P_{s,0} &= \frac{\beta }{\gamma}, \\
P_{s,1} &= \frac{1-\beta}{1-\gamma}.
\end{align*}
On the other hand, the relay transmission only affects the destination output for a fraction $(1-\gamma)$ of the time, i.e., when $S_r=1$, hence the relay must exploit all its available power when $S_r=1$; we therefore let
\begin{align*}
P_{r,0} &= 0,  \\
P_{r,1} &= \frac{1}{1-\gamma}.
\end{align*}
With this, the cut-set upper bound $r^{\rm(CS-HD)}$ in~\eqref{eq:and finally N-max-h} can be rewritten as \eqref{eq:eqnewouter} at the top of next page
\begin{figure*}
\begin{align}
  r^{\rm(CS-HD)} 
  &= \max_{(\gamma, |\alpha_1|, \beta)\in[0,1]^3} \min \Big\{ \left. \mathcal{H}(\gamma) +\gamma \log \left (1+\frac{S  \beta}{\gamma} \right )+(1-\gamma)\log \left(1+\frac{I}{1-\gamma}+\frac{S(1-\beta)}{1-\gamma}\nonumber +2|\alpha_1|\sqrt{\frac{I}{1-\gamma}\frac{S(1-\beta)}{1-\gamma} }\right), \right.
\\& \left. \qquad \qquad \qquad \qquad \qquad 0+\gamma \log \left (1+\frac{C\beta}{\gamma}+\frac{S\beta}{\gamma} \right)+(1-\gamma)\log\left (1+(1-|\alpha_1|^2) \frac{S(1-\beta)}{1-\gamma} \right) \right\}\nonumber
\\& \leq \max_{\gamma \in [0,1]} \min \Big\{\left. \mathcal{H}(\gamma) +\gamma \log \left (1+\frac{S}{\gamma} \right )   \nonumber
 +(1-\gamma)\log \left(1+\left(\sqrt{\frac{I}{1-\gamma}}+\sqrt{\frac{S}{1-\gamma}}\right)^2\right), \right.
\\& \left. \qquad \qquad \qquad 0+\gamma \log \left (1+\frac{C}{\gamma}+\frac{S}{\gamma} \right)+(1-\gamma)\log\left (1+\frac{S}{1-\gamma} \right) \right\}\nonumber
\\& = \max_{\gamma \in [0,1]} \min \Big\{ 2\mathcal{H}(\gamma)+ \gamma \log \left (\gamma+S \right )   \nonumber
 +(1-\gamma)\log \left(1-\gamma+\left(\sqrt{I}+\sqrt{S}\right)^2\right), 
\\& \left. \qquad \qquad \qquad \mathcal{H}(\gamma)+\gamma \log \left (\gamma+C+S \right)+(1-\gamma)\log\left (1-\gamma+S \right) \right \}\nonumber
\\& \leq 2\log(2) + \max_{\gamma \in [0,1]} \min \Big\{ \gamma \log \left (1+S \right )  \nonumber 
 +(1-\gamma)\log \left(1+\left(\sqrt{I}+\sqrt{S}\right)^2\right), \left.  \gamma \log \left (1+C+S \right)+(1-\gamma)\log\left (1+S \right) \right \}\nonumber
\\&= 2\log(2)+ \log \left (1+S \right ) \max_{\gamma \in [0,1]} \min \left \{\nonumber
\gamma +(1-\gamma) b_1,
\gamma b_2 +(1-\gamma)\right \} 
\\&= 2\log(2)+ \log \left (1+S \right ) \left( 1 + \max_{\gamma \in [0,1]} \min \left \{\nonumber
(1-\gamma) (b_1-1),
\gamma (b_2-1)\right \}  \right)
\\& = 2\log(2)+ \log \left (1+S \right )\left(1+\frac{(b_1 -1)(b_2-1)}{(b_1-1) + (b_2-1)}\right),
\label{eq:eqnewouter}
\end{align}
\end{figure*}
where we defined $b_1$ and $b_2$ as in~\eqref{eq:def of b1}-\eqref{eq:def of b2}, namely
\begin{align*}
  & b_1 := \frac{\log \left(1+(\sqrt{I}+\sqrt{S})^2\right)}{\log \left (1+S \right )} > 1 \ \text{since $I>0$},
\\& b_2 := \frac{\log \left (1+C+S \right)}{\log \left (1+S \right )} > 1 \ \text{since $C>0$}.
\end{align*}
Note that the optimal $\gamma$ is found by equating the two arguments of the $\max \min$ and is given by
\begin{align*}
\gamma^{*}_{\rm CS} := \frac{(b_1-1)}{(b_1-1) + (b_2-1)}. 
\end{align*}
This proves the upper bound in~\eqref{eq:cuset for gap analytical}.
\end{itemize}
%
\end{IEEEproof}

\section{Proof of Proposition~\ref{prop:cutset dof}}
\label{app: proof of prop:cutset dof}
\begin{IEEEproof}
The upper bound in~\eqref{eq:cuset for gap analytical} implies
\begin{align*}
  &\mathsf{d}^{\rm(HD-RC)}
 \\ & \leq 
\lim_{\mathsf{SNR}\to+\infty}\frac{\log \left (1+S \right )}{\log \left (1+\mathsf{SNR} \right )}\left(1+\frac{(b_1-1)(b_2-1)}{(b_1-1) + (b_2-1)}\right)
\\& = \beta_{\rm sd}\left(1+\frac{[\beta_{\rm rd}/\beta_{\rm sd} -1]^+ \ [\beta_{\rm sr}/\beta_{\rm sd} -1]^+}{[\beta_{\rm rd}/\beta_{\rm sd} -1]^+ + [\beta_{\rm sr}/\beta_{\rm sd} -1]^+} \right)
\\& = \beta_{\rm sd} + \frac{[\beta_{\rm rd}-\beta_{\rm sd}]^+ \ [\beta_{\rm sr}-\beta_{\rm sd}]^+}{[\beta_{\rm rd}-\beta_{\rm sd}]^+ + [\beta_{\rm sr}-\beta_{\rm sd}]^+},
\end{align*}
since $b_1 \to \max\{\beta_{\rm sd},\beta_{\rm rd}\}/\beta_{\rm sd}$ and $b_2 \to \max\{\beta_{\rm sd},\beta_{\rm sr}\}/\beta_{\rm sd}$ at high $\mathsf{SNR}$,
which is equivalent to the right hand side of~\eqref{eq:dofHDrelay} after straightforward manipulations.
\end{IEEEproof}

\section{Proof of Proposition~\ref{prop:inner upper bounds}}
\label{app: proof of prop:inner upper bounds}

The largest achievable rate for the memoryless relay channel is the combination of PDF and CF/NNC proposed in the seminal work of Cover and ElGamal~\cite{coveElGamal}.  
Here we use PDF. The bound in~\eqref{eq:inner for gap pdf fixed and numerical} follows since:
\begin{IEEEproof}
The PDF scheme in~\cite[Thm.{16.3}]{book:ElGamalKim2012} adapted to the HD model gives the following rate lower bound 
\begin{align*}
&C^{\rm(HD-RC)}
\\&\geq \max_{P_{U,X_s,X_r,S_r}}
\min\Big\{
I(S_r;Y_d)
+I(X_s,X_r; Y_d|S_r),
\\& \quad \quad \quad \quad \quad \quad \quad \quad I(U; Y_r| X_r,S_r)+I(X_s; Y_d|U,X_r,S_r)
\Big\}
\\&\geq
\max\min\Big\{ I_0^{\rm(PDF)}+
 \gamma I_{5} +(1-\gamma)I_{6},
 \gamma I_{7} +(1-\gamma)I_{8}
\Big\}
\\&= r^{\rm(PDF-HD)} \ \text{in~\eqref{eq:inner for gap pdf fixed and numerical}}, 
\end{align*}
where for the last inequality we let $\gamma:=\mathbb{P}[S_r=0]\in[0,1]$ be the fraction of time the relay listens and, conditioned on $S_r=\ell$, $\ell\in\{0,1\}$, we consider the following jointly Gaussian input
\begin{align*}
&\left.
\begin{pmatrix}
U \\ 
\frac{X_s}{\sqrt{P_{s,\ell}}} \\
\frac{X_r}{\sqrt{P_{r,\ell}}} \\ 
\end{pmatrix}\right|_{S_r=\ell}
\sim \mathcal{N}\left( \mathbf{0},
\begin{bmatrix}
1 & \rho_{t|\ell} & \rho_{r|\ell} \\
\rho_{t|\ell}^* & 1 & \alpha_{\ell} \\ 
\rho_{r|\ell}^* & \alpha_{\ell}^* & 1\\
\end{bmatrix}
\right) 
\\& \quad \quad \quad \quad \quad \quad \quad \quad \quad:
\begin{bmatrix}
1 & \rho_{t|\ell} & \rho_{r|\ell} \\
\rho_{t|\ell}^* & 1 & \alpha_{\ell} \\ 
\rho_{r|\ell}^* & \alpha_{\ell}^* & 1\\
\end{bmatrix}
\succeq \mathbf{0}.
\end{align*}
In particular, we use specific values for the parameters $\{\rho_{t|\ell},\rho_{r|\ell},\alpha_{\ell}\}_{\ell\in\{0,1\}}$, namely
\begin{subequations}
\begin{align}
  &\angle{\alpha_{1}} + \theta =0, \label{eq:assumption a1}
\\&\alpha_{0} = 0 \ \text{and either} \ |\rho_{t|0}|^2=1-|\rho_{r|0}|^2=0 \nonumber
\\& \quad \quad \quad \quad \text{or} \ |\rho_{r|0}|^2=1-|\rho_{t|0}|^2=0, \label{eq:assumption state0}
\\&\rho_{t|1}= \alpha_{1}^*, \ \rho_{r|1}=1. \label{eq:assumption state1}
\end{align}
\end{subequations}
With these definitions, the mutual information terms $I_0^{\rm(PDF)},I_{5},\ldots,I_8$ in~\eqref{eq:inner for gap pdf fixed and numerical} are 
\begin{align*}
   &I(X_s,X_r; Y_d|S_r=0) = \log\left(1+S P_{s,0}\right) =: I_{5}; 
\\& I(X_s,X_r; Y_d|S_r=1) 
\\&= \log\left(1+S P_{s,1}+I P_{r,1} +2|\alpha_1|\sqrt{  S P_{s,1} \ IP_{r,1}}\right)=: I_{6},
\end{align*}
(note $I_{5}=I_{1}$ and $I_{6} = I_{2}$ because of the assumption in~\eqref{eq:assumption a1});
next, by using the assumption in~\eqref{eq:assumption state0}, that is, in state $S_r=0$ the inputs $X_s$ and $X_r$ are independent, and that either $U=X_s$ or $U=X_r$, we have: 
if $U=X_s$ independent of $X_r$
\begin{figure*}
\begin{align}
  r^{\rm(PDF-HD)}
  &= \max_{\gamma \in [0,1], |\alpha|\leq 1, \beta\in[0,1]}\min \left \{ I_0^{\rm(PDF)}+ \right.\left.  \gamma \log \left ( 1+\frac{\beta S}{\gamma}\right) \right. \nonumber
  \\& \left. \qquad \qquad \qquad \qquad \qquad +(1-\gamma)\log \left( 1+\frac{S (1-\beta) }{1-\gamma}+\frac{I}{1-\gamma}+2|\alpha|\sqrt{\frac{S(1-\beta) }{1-\gamma}\frac{I}{1-\gamma}}  \right), \right. \nonumber
\\& \left. \qquad \qquad \qquad \qquad \qquad \gamma \log \left ( 1+\frac{1}{\gamma}\max \left \{ C\beta ,S\beta  \right\} \right)+(1-\gamma)\log \left( 1+(1-|\alpha|^2) \frac{S(1-\beta) }{1-\gamma} \right)\right \} \nonumber
\\&\geq  \max_{\gamma \in [0,1],\beta\in[0,1]}\min \left \{ 0+\gamma \log \left ( 1+\frac{\beta S}{\gamma}\right)+(1-\gamma)\log \left( 1+\frac{S(1-\beta) }{(1-\gamma)}+\frac{I}{1-\gamma} \right), \right.\nonumber
\\& \qquad \qquad \qquad \qquad \left. \gamma \log \left ( 1+\frac{1}{\gamma}\max \left \{ \beta C,\beta S \right\} \right)+(1-\gamma)\log \left( 1+ \frac{S(1-\beta) }{(1-\gamma)} \right)\right \}  \nonumber
\\& \geq \max_{\gamma \in [0,1]} \min \left \{ \gamma \log \left( 1+S \right)+(1-\gamma)\log \left ( 1+S+I \right), \right.\left.  \gamma \log \left( 1+\max \left \{ C,S\right\} \right)+(1-\gamma)\log \left( 1+S \right) \right \}\nonumber
\\&= \log \left (1+S \right ) \max_{\gamma \in [0,1]} \min \left \{\nonumber
\gamma +(1-\gamma) c_1,
\gamma c_2 +(1-\gamma)\right \} 
\\&= \log \left (1+S \right ) \left( 1 + \max_{\gamma \in [0,1]} \min \left \{\nonumber
(1-\gamma) (c_1-1),
\gamma (c_2-1)\right \}  \right)
\\& = \log \left (1+S \right )\left(1+\frac{(c_1 -1)(c_2-1)}{(c_1-1) + (c_2-1)}\right),
\label{eq:eqnewinner}
\end{align}
\end{figure*}
\begin{align*}
  &I(U; Y_r| X_r,S_r=0) + I(X_s; Y_d|U,X_r,S_r=0) = 
\\&=I(X_s; \sqrt{C} X_s + Z_r| X_r,S_r=0)
\\&+ I(X_s; \sqrt{S} X_s + Z_d|X_s,X_r,S_r=0)
\\&=\log\left(1+C P_{s,0}\right) +0,
\end{align*}
and if $U=X_r$ independent of $X_s$
\begin{align*}
  &I(U; Y_r| X_r,S_r=0) + I(X_s; Y_d|U,X_r,S_r=0) = 
\\&=I(X_r; \sqrt{C} X_s + Z_r| X_r,S_r=0) 
\\&+ I(X_s; \sqrt{S} X_s + Z_d|X_r,S_r=0)
\\&=0+\log\left(1+S P_{s,0}\right);
\end{align*}
therefore under the assumption in~\eqref{eq:assumption state0} we have
\begin{align*}
  &I(U; Y_r| X_r,S_r=0) + I(X_s; Y_d|U,X_r,S_r=0) 
  \\&= \log\left(1+\max\{C,S \}P_{s,0}\right) =: I_7;
\end{align*}
next, by using the assumption in~\eqref{eq:assumption state1}, that is, in state $S_r=1$ we let $U=X_r$, we have 
\begin{align*}
  & I(U; Y_r| X_r,S_r=1)+I(X_s; Y_d|U,X_r,S_r=1)
\\&=I(X_r; Z_r| X_r,S_r=1)+I(X_s;\sqrt{S} X_s   + Z_d|X_r,S_r=1) 
\\&=0+I(X_s;\sqrt{S} X_s + Z_d|X_r,S_r=1) 
\\&= \log\left( 1+S (1-|\alpha_{1}|^2) P_{s,1}\right) =: I_8, 
\end{align*}
(note $I_{7}\leq I_{3}$ and $I_{8} = I_{4}$);
finally 
\begin{align*}
  &I(S_r;Y_d)
  \\&=\mathbb{E}\left[\log\frac{1}{f_{Y_d}(Y_d)}\right]
   \!-\![\gamma \log(v_0) \!+\!(1-\gamma)\log(v_1)+\log(\pi {\rm e})] 
   \\&=: I_0^{\rm(PDF)},
\end{align*}
where $f_{Y_d}(\cdot)$ is the density of the destination output $Y_d$, which is a mixture of (proper complex) Gaussian random variables, i.e., 
\begin{align*}
 f_{Y_d}(t) &= \frac{  \gamma}{\pi v_0} \exp(-|t|^2/v_0) + \frac{1-\gamma}{\pi v_1} \exp(-|t|^2/v_1), \ t\in\mathbb{C},
\\& v_0 := {\rm Var}[Y_d|S_r=0] =  \exp(I_5), 
\\& v_1 := {\rm Var}[Y_d|S_r=1] =  \exp(I_6). 
\end{align*}
Note that $I_0^{\rm(PDF)} = I(S_r;Y_d) \leq H(S_r) = \mathcal{H}(\gamma)$.
This proves the lower bound in~\eqref{eq:inner for gap pdf fixed and numerical}.

Next we show how to further lower bound the rate in~\eqref{eq:inner for gap pdf fixed and numerical} to obtain the rate expression in~\eqref{eq:inner for gap analytical}. With the same parameterization of the powers as in Appendix~\ref{app: proof of prop:cutset upper bounds}, namely
\begin{align*}
P_{s,0} &= \frac{\beta }{\gamma}, \\
P_{s,1} &= \frac{1-\beta}{1-\gamma}, \\
P_{r,0} &= 0,  \\
P_{r,1} &= \frac{1}{1-\gamma}.
\end{align*}
we have \eqref{eq:eqnewinner} at the top of this page
where we defined $c_1$ and $c_2$ as in~\eqref{eq:def of c1}-\eqref{eq:def of c2}, namely
\begin{align*}
 c_1 &:= \frac{\log \left(1+I+S\right)}{\log \left (1+S \right )} \geq 1 \ \text{since $I>0$}, 
 \\
 c_2 &:= \frac{\log \left(1+\max\{C,S\} \right)}{\log \left (1+S \right )}\geq 1 \ \text{since $C>0$}. 
\end{align*}
Notice that $c_i \leq b_i, i=1,2,$ where $b_i, i=1,2,$ are defined in~\eqref{eq:def of b1}-\eqref{eq:def of b2}.
The optimal $\gamma$, indicated by $\gamma_{\rm PDF}^*$ is given by
\begin{align*}
\gamma_{\rm PDF}^* := \frac{(c_1-1)}{(c_1-1) + (c_2-1)} \ \in \ [0,1].
\end{align*}
%
\end{IEEEproof}

\begin{figure*}
\begin{align}
C^{\rm(HD-RC)} \nonumber
&\geq
 \max_{P_{Q}P_{X_s|Q}P_{[X_r,S_r]|Q}P_{\widehat{Y}_r|[X_r,S_r],Y_r,Q} : |Q|\leq 2} \min\Big\{
    I(X_s; \widehat{Y}_r,Y_d|[X_r,S_r],Q),
\\& \qquad I(X_s,[X_r,S_r]; Y_d|Q)-I(Y_r; \widehat{Y}_r| X_s,[X_r,S_r],Y_d,Q)
\Big\}\nonumber
\\& = \max_{P_{Q}P_{S_r|Q} P_{X_s|Q}P_{X_r|S_r,Q}P_{\widehat{Y}_r|X_r,Y_r,S_r,Q} : |Q|\leq 2} \min\Big\{
    I(X_s; \widehat{Y}_r,Y_d|Q,S_r,X_r),\nonumber
\\& \qquad I(S_r; Y_d|Q) + I(X_s,X_r; Y_d|S_r, Q)-I(Y_r; \widehat{Y}_r| X_s,X_r,Y_d,S_r,Q)\nonumber
\Big\}
\\&\geq r^{\rm(NNC-HD)} \ \text{in~\eqref{eq:GHDRC NNC rate}}, 
\label{eq:eqnewNNC}
\end{align}
\end{figure*}

\begin{remark}
\label{rem:simple pdf}
A further lower bound on the PDF rate $r^{\rm(PDF-HD)}$ in~\eqref{eq:inner for gap pdf fixed and numerical} can be obtained by trivially lower bounding $I_0^{\rm(PDF)}\geq 0$, which corresponds to a fixed transmit/receive schedule for the relay. 
\end{remark}

\section{Proof of Proposition~\ref{prop:inner dof}}
\label{app: proof of prop:inner dof}

\begin{IEEEproof}
The lower bound in~\eqref{eq:inner for gap analytical} implies
\begin{align*}
  \mathsf{d}^{\rm(HD-RC)} 
  & \!\geq \!
\lim_{S\to+\infty}\frac{\log \left (1+S \right )}{\log \left (1\!\!+\!\!\mathsf{SNR} \right )} \left( 1+\frac{(c_1-1)( c_2 - 1)}{\left( c_1 -1 \right)+ \left( c_2 -1 \right)} 
\right )
\\& = \beta_{\rm sd}\left(1+\frac{[\beta_{\rm rd}/\beta_{\rm sd} -1]^+ \ [\beta_{\rm sr}/\beta_{\rm sd} -1]^+}{[\beta_{\rm rd}/\beta_{\rm sd} -1]^+ + [\beta_{\rm sr}/\beta_{\rm sd} -1]^+} 
\right)
\\& = \beta_{\rm sd} + \frac{[\beta_{\rm rd}-\beta_{\rm sd}]^+ \ [\beta_{\rm sr}-\beta_{\rm sd}]^+}{[\beta_{\rm rd}-\beta_{\rm sd}]^+ + [\beta_{\rm sr}-\beta_{\rm sd}]^+},
\end{align*}
since $c_1 \to \max\{\beta_{\rm sd},\beta_{\rm rd}\}/\beta_{\rm sd}$ and $c_2 \to \max\{\beta_{\rm sd},\beta_{\rm sr}\}/\beta_{\rm sd}$ at high $\mathsf{SNR}$, which is equivalent to the right hand side of~\eqref{eq:dofHDrelay} after straightforward manipulations.
\end{IEEEproof}

\section{Proof of Proposition~\ref{prop:lda}}
\label{app: proof of prop:inner dof LDA}

\begin{IEEEproof}
The rate in~\eqref{eq:lda inner bound} can be further lower bounded as
\begin{align*}
r^{\rm(LDA-HD)} \!\geq \! -\log(2)\!+\!\log \left (1\!+\!S \right )\left(1\!+\!\frac{(c_3 \!-\!1)(c_4\!-\!1)}{(c_3\!-\!1) + (c_4\!-\!1)}\right),
\end{align*}
where $c_3\!:=\!c_1\!=\!\frac{\log \left( 1+I+S\right)}{\log \left( 1+S\right)}$ and $c_4\!:=\!b_2\!=\!\frac{\log \left( 1+C+S\right)}{\log \left( 1+S\right)}$.
The rate above implies
\begin{align*}
  \mathsf{d}
  & \geq 
\lim_{S\to+\infty}\frac{\log \left (1+S \right )}{\log \left (1+\mathsf{SNR} \right )} \left( 1+\frac{(c_3-1)( c_4 - 1)}{\left( c_3 -1 \right)+ \left( c_4 -1 \right)} 
\right )
\\& = \beta_{\rm sd}\left(1+\frac{[\beta_{\rm rd}/\beta_{\rm sd} -1]^+ \ [\beta_{\rm sr}/\beta_{\rm sd} -1]^+}{[\beta_{\rm rd}/\beta_{\rm sd} -1]^+ + [\beta_{\rm sr}/\beta_{\rm sd} -1]^+} 
\right)
\\& = \beta_{\rm sd} + \frac{[\beta_{\rm rd}-\beta_{\rm sd}]^+ \ [\beta_{\rm sr}-\beta_{\rm sd}]^+}{[\beta_{\rm rd}-\beta_{\rm sd}]^+ + [\beta_{\rm sr}-\beta_{\rm sd}]^+},
\end{align*}
since $c_3 \to \max\{\beta_{\rm sd},\beta_{\rm rd}\}/\beta_{\rm sd}$ and $c_4 \to \max\{\beta_{\rm sd},\beta_{\rm sr}\}/\beta_{\rm sd}$ at high $\mathsf{SNR}$, which is equivalent to the right hand side of~\eqref{eq:dofHDrelay} after straightforward manipulations.
\end{IEEEproof}

\section{Achievable rate with NNC}
\label{app: inner NNC bounds}
The largest achievable rate for the memoryless relay channel is the combination of PDF and CF/NNC proposed in the seminal work of Cover and ElGamal~\cite{coveElGamal}.  
Here we use NNC to show:
\begin{proposition}
\label{prop:NNC-based Strategy for the G-HD-RC}
The capacity of the G-HD-RC is lower bounded as
\begin{subequations}
\begin{align}
&C^{\rm(HD-RC)}
\geq  r^{\rm(NNC-HD)} \nonumber
\\&:= 
\max\min\Big\{
I_{0}^{\rm(NNC)}+\!\!\!\! \sum_{(i,j)\in[0:1]^2} \!\!\!\! \!\!\gamma_{ij} I_{9,ij}, 
                  \!\!\!\! \sum_{(i,j)\in[0:1]^2} \!\!\!\! \!\!\gamma_{ij} I_{10,ij}
\Big\},
\label{eq:GHDRC NNC rate}
\end{align}
where the maximization is over
\begin{align}
  & \gamma_{ij}\in[0,1] : \sum_{(i,j)\in[0:1]^2} \gamma_{ij} = 1, \label{eq:GHDRC NNCgamma}
\\& P_{s,i}  \geq 0 : \sum_{(i,j)\in[0:1]^2} \gamma_{ij} \ P_{s,i}  \leq 1, \label{eq:GHDRC NNCpower relay} 
\\& P_{r,ij} \geq 0 : \sum_{(i,j)\in[0:1]^2} \gamma_{ij} \ P_{r,ij} \leq 1, \label{eq:GHDRC NNCpower source}
\end{align}
\label{eq:GHDRC NNC}
\end{subequations}
where the different mutual information terms in~\eqref{eq:GHDRC NNC} are defined next.
\end{proposition}

\begin{IEEEproof}
The NNC scheme in~\cite[Remark {18.5}]{book:ElGamalKim2012} adapted to the HD model gives the rate lower bound in \eqref{eq:eqnewNNC} at the top of this page, where the mutual information terms $\{I_{9,ij},I_{10,ij}\}$, $(i,j)\in[0:1]^2$ and $I_{0}^{\rm(NNC)}$ in~\eqref{eq:GHDRC NNC rate} are obtained as follows. We consider the following assignment on the inputs and on the auxiliary random variables for each $(i,j)\in[0:1]^2$
\begin{align*}
  &\mathbb{P}[Q=i,S_r=j] = \gamma_{ij} \ \text{such that~\eqref{eq:GHDRC NNCgamma} is satisfied},
\\&\left.\begin{pmatrix}
X_s \\ X_r \\
\end{pmatrix}\right|_{Q=i,S_r=j}
\sim \mathcal{N}\left( \mathbf{0},
\begin{bmatrix}
P_{s,i} & 0 \\ 
0 & P_{r,ij}\\
\end{bmatrix}
\right) 
\\& \text{such that~\eqref{eq:GHDRC NNCpower relay} and~\eqref{eq:GHDRC NNCpower source} are satisfied},
\\&\widehat{Y}_r|_{X_r,Y_r,Q=i,S_r=j} = Y_r + \widehat{Z}_{r,ij},
\\& \text{$\widehat{Z}_{r,ij}\sim\mathcal{N}(0,\sigma_{ij}^2)$ and independent of everything else},
\end{align*}
and in order to meet the constraint that $X_s$ cannot depend on $S_r$ conditioned on $Q$ we must impose the constraint that in state $Q=i,S_r=j$ the power of the source only depends on the index $i$. Then for each $(i,j)\in[0:1]^2$
\begin{align}
&I(X_s; \widehat{Y}_r,Y_d|X_r,Q=i,S_r=j) \nonumber
\\& = 
\log\left(1+\left(S+\frac{C(1-j)}{1+\sigma^2_{ij}}\right)P_{s,i}\right)
=: I_{10,ij}, \label{eq:firstnnc}
\\
&I(X_s,X_r; Y_d|Q=i,S_r=j)+ \nonumber
\\&-I(Y_r; \widehat{Y}_r| X_s,X_r,Y_d,Q=i,S_r=j) \nonumber
\\& = \log\left(1\!+\!SP_{s,i}\!+\!I  j  P_{r,ij}\right)\!
   -\! \log\left(1\!+\!\frac{1}{\sigma^2_{ij}}\right)
=: I_{9,ij}, \label{eq:secondnnc}
\\&I(S_r;Y_d|Q) = -\sum_{(i,j)}\gamma_{ij} \log(v_{ij})-\log(\pi {\rm e})\nonumber
\\&\quad \nonumber
   +(\gamma_{00}+\gamma_{01}) \ \mathbb{E}\left[\log\frac{1}{f_0(Y)} | Q=0\right]
 \\& \quad  +(\gamma_{10}+\gamma_{11}) \ \mathbb{E}\left[\log\frac{1}{f_1(Y)} | Q=1\right] \nonumber
=: I_{0}^{\rm(NNC)}
\\&Y_d|_{Q=0} \sim f_0(t) := \frac{\gamma_{00}}{\gamma_{00}+\gamma_{01}} \ \frac{1}{\pi v_{00}} \exp(-|t|^2/v_{00}) \nonumber
\\& \quad \quad \quad \quad \quad \quad  \quad        + \frac{\gamma_{01}}{\gamma_{00}+\gamma_{01}} \ \frac{1}{\pi v_{01}} \exp(-|t|^2/v_{01}), \ t\in\mathbb{C},\nonumber
\\&Y_d|_{Q=1} \sim f_1(t) := \frac{\gamma_{10}}{\gamma_{10}+\gamma_{11}} \ \frac{1}{\pi v_{10}} \exp(-|t|^2/v_{10})\nonumber
\\& \quad \quad \quad \quad \quad \quad  \quad           + \frac{\gamma_{11}}{\gamma_{10}+\gamma_{11}} \ \frac{1}{\pi v_{11}} \exp(-|t|^2/v_{11}), \ t\in\mathbb{C},\nonumber
\\& v_{ij} := {\rm Var}[Y_d|Q=i,S_r=j] = 1+S \ P_{s,i}+I \ j \ P_{r,ij}.\nonumber
\end{align}
This proves the lower bound in~\eqref{eq:GHDRC NNC} as a function of $\sigma_{ij}^2, (i,j)\in\{0,1\}^2$.

In order to find the optimal $\sigma_{ij}^2, (i,j)\in\{0,1\}^2$ we reason as follows.
$I_{10,ij}$ in \eqref{eq:firstnnc} is decreasing in $\sigma_{ij}^2$ while $I_{9,ij}$ in \eqref{eq:secondnnc} is increasing. At the optimal point these two rates are the same. Let
\begin{align*}
C_i &:= 1+\frac{C P_{s,i}}{1+S P_{s,i}}, \\ 
x_i &:= \frac{1}{\sigma_{i0}^2}, \\
I^\prime &:= I(S_r,X_r; Y_d| Q),
\end{align*}
and rewrite the lower bound in~\eqref{eq:GHDRC NNC} as
\begin{align*}
&r^{\rm(NNC-HD)} 
=(\gamma_{00}+\gamma_{01}) \log(1+S P_{s,0})
\\& \quad \quad +(\gamma_{10}+\gamma_{11}) \log(1+S P_{s,1})
\\&
\quad \quad- \gamma_{00}\log\left(1+x_0\right)
- \gamma_{10}\log\left(1+x_1\right)
\\&
\quad \quad+\min\Big\{ \gamma_{00}\log\left(1\!+\!x_0 C_{0}\right)
           +\gamma_{10}\log\left(1\!+\!x_1 C_{1}\right),
           I^\prime
           \Big\}.
\end{align*}
The solution of
\begin{align*}
&\min_{(x_0,x_1)\in\mathbb{R}^2_+}\Big\{
 \gamma_{00}\log\left(1+x_0\right)
+\gamma_{10}\log\left(1+x_1\right)
\Big\}
\\&\text{subject to}
\quad
 \gamma_{00}\log\left(1+ x_0 C_{0} \right)
+\gamma_{10}\log\left(1+ x_1 C_{1} \right) = I^\prime
\end{align*}
can be found to be
\begin{align*}
x_i = \frac{[\eta C_i - 1]^+}{(1-\eta)C_i}, \ i\in\{1,2\},
\end{align*}
with $\eta\leq 1$ such that
\begin{align*}
\gamma_{00}\log\left(1+ x_0 C_{0} \right)
+\gamma_{10}\log\left(1+ x_1 C_{1} \right) = I^\prime.
\end{align*}
\end{IEEEproof}

\begin{figure*}
\begin{align}
  &r^{\rm(NNC-HD)} \nonumber
   \geq \max_{\gamma \in [0,1], \sigma_0^2\geq 0, \beta\in[0,1]}
    \min \left \{ \gamma \log \left ( 1+\frac{\beta S}{\gamma}\right)-\gamma \log \left( 1+\frac{1}{\sigma^2_0} \right)+\right.
\\& \qquad \left. +(1-\gamma)\log \left( 1+\frac{(1-\beta)S}{1-\gamma}+\frac{I}{1-\gamma} \right), \right. \nonumber
\\& \qquad \left.  \gamma \log \left ( 1+\frac{C\beta}{(1+\sigma^2_0)\gamma}+\frac{S \beta}{\gamma}\right)+(1-\gamma)\log \left( 1+\frac{(1-\beta)S}{1-\gamma} \right) \right \} \nonumber
\\&\stackrel{\beta=\gamma}{\geq} \max_{\gamma \in [0,1], \sigma_0^2\geq 0}\nonumber
   \min \left \{ \gamma \log \left ( 1+S\right)+(1-\gamma)\log \left( 1+S+I \right), \right. 
\\& \qquad  \left.  \gamma \log \left ( 1+\frac{C}{1+\sigma_0^2}+S\right)+(1-\gamma)\log \left(1+S \right) \right \}\nonumber
 - \gamma \log \left( 1+\frac{1}{\sigma^2_0} \right)
\\&= \max_{\gamma \in [0,1], \sigma_0^2\geq 0}\left[\nonumber
\log \left ( 1+S\right)  
   \min  \left \{ \gamma +(1-\gamma)c_5, 
                  \gamma c_6+(1-\gamma) \right \}
 - \gamma \log \left( 1+\frac{1}{\sigma^2_0} \right)
 \right]
\\& \stackrel{\gamma=\gamma_{\rm NNC}^*}{\geq}  \max_{\sigma_0^2\geq 0}\nonumber
\log \left( 1+S \right)\left(1+\frac{(c_5-1) (c_6-1)}{(c_5-1) + (c_6-1)} \left(1-\frac{\log \left( 1+\frac{1}{\sigma^2_0} \right)}{\log \left(1+\frac{C}{(1+\sigma_0^2)(1+S)} \right)}\right)\right) 
\\& \stackrel{\sigma_0^2=1}{\geq}  -\log(2) + 
\log \left( 1+S \right)\left(1+\frac{(c_5-1) (c_6-1)}{(c_5-1) + (c_6-1)} \right),
\label{eq:eqnewlowNNC}
\end{align}
\end{figure*}
\begin{remark}
\label{rem:simple nnc}
For the special case of $Q=S_r$, that is, $I_0^{\rm(NNC)}=I(S_r;Y_d|Q)=I(Q;Y_d|Q)=0$, the achievable rate in Proposition~\ref{prop:NNC-based Strategy for the G-HD-RC} reduces to\begin{subequations}
\begin{align}
&r^{\rm(NNC-HD)} \geq
\max_{(\gamma,\beta)\in[0,1]^2}\min\Big\{
 \gamma I_{9}  +(1-\gamma)I_{10}, \nonumber
 \\& \quad \quad \quad \quad \quad \quad \quad \gamma I_{11} +(1-\gamma)I_{12}
\Big\}, \label{eq:GHDRC NNCsimplified rate}
\\&I_{9}  := \log\left(1+S P_{s,0}\right)-\log \left ( 1+\frac{1}{\sigma_{0}^2} \right ),
\\&I_{10} := \log\left(1+S P_{s,1}+I P_{r,1} \right),
\\&I_{11} := \log\left(1+S P_{s,0}+\frac{C }{1+\sigma_{0}^2} P_{s,0} \right)
\\&I_{12}:= \log\left( 1+ S P_{s,1} \right).
\\ & \sigma_{0}^2 := \frac{B+1}{(1+A)^{\frac{1}{\gamma}-1}-1}, \label{eq:GHDRC NNCsimplified sigma0} 
\\ & A := \frac{I P_{r,1}}{1+ S P_{s,1}}, \quad B := \frac{C P_{s,0}}{1+ S P_{s,0}},
\\ & P_{s,0} = \frac{\beta}{\gamma},\quad  P_{s,1} = \frac{1-\beta}{1-\gamma},\quad P_{r,1} = \frac{1}{1-\gamma},
\end{align}
\label{eq:GHDRC NNCsimplified}
\end{subequations}
where the optimal value for $\sigma_{0}^2$ in~\eqref{eq:GHDRC NNCsimplified sigma0} is obtained by equating the two expressions within the $\min$ in~\eqref{eq:GHDRC NNCsimplified rate}. 
\end{remark}

\begin{proposition}
NNC with deterministic switch achieves the gDoF upper bound in~\eqref{eq:dofHDrelay}.
\end{proposition} 

\begin{IEEEproof}
With the achievable rate in Remark~\ref{rem:simple nnc} (where here we explicitly write the optimization wrt $\sigma_0^2$) we have \eqref{eq:eqnewlowNNC} at the top of next page,
where we defined $c_5$ and $c_6$ as
\begin{align*}
c_5&=c_1 := \frac{\log \left(1+I+S\right)}{\log \left (1+S \right )} \geq 1  \ \text{since $I>0$ and as in~\eqref{eq:def of c1}},
\\
c_6 &:= \frac{\log \left(1+\frac{C}{1+\sigma_0^2}+S \right)}{\log \left (1+S \right )}\geq 1  \ \text{since $C>0$},
\end{align*}
and where 
\begin{align*}
\gamma_{\rm NNC}^* := \frac{(c_5-1)}{(c_5-1) + (c_6-1)} \ \in \ [0,1].
\end{align*}
By reasoning as for the PDF in Appendix~\ref{app: proof of prop:inner dof}, it follows from the last rate bound that NNC also achieves the gDoF in~\eqref{eq:dofHDrelay}.

\end{IEEEproof}

\begin{remark}
\label{rem:simple nnc 2}
For the special case of $Q=\emptyset$, i.e.,  the time-sharing variable $Q$ is a constant, the achievable rate in Proposition~\ref{prop:NNC-based Strategy for the G-HD-RC} reduces to
\begin{align*}
&r^{\rm(NNC-HD)} \geq \!\!\max_{P_{X_s}P_{X_r,S_r}P_{\widehat{Y}_r|[X_r,S_r],Y_r}} 
\!\!\!\!\!\!\!\!\!\!\min\Big\{
    I\left(X_s; \widehat{Y}_r,Y_d| S_r,X_r\right)\!\!,
\\& I\left(X_s,X_r,S_r; Y_d\right)
   \!-\!I\left(Y_r; \widehat{Y}_r| S_r,X_r, X_s,Y_d\right)
\Big\}
\\& \geq \!\!\!\max_{\gamma\in[0,1], \sigma^2} \!\!\!\! \min \left \{ 
\gamma \log \left( 1\!+\!S\!+\!\frac{C}{1+\sigma^2} \right)\!+\!(1\!-\!\gamma) \log \left( 1\!+\!S \right), \right.
\\& \left. 
I\left(S_r; Y_d\right) +
\gamma \log \left( 1+S\right) -\gamma \log \left( 1+\frac{1}{\sigma^2} \right) \right.
\\& \left. +(1-\gamma) \log \left( 1+S+\frac{I}{1-\gamma} \right)\right \}.
\end{align*}
Note that with $Q=\emptyset$ the source always transmits with constant power, regardless of the state of the relay, while the relay sends only when in transmitting mode. Thus in this particular setting there is no coordination between the source and the relay.  
\end{remark}

%

\section{Proof of Proposition~\ref{prop:gap pdf rand}}\label{app:prop:gap pdf rand}
\begin{IEEEproof}
Consider the upper bound in~\eqref{eq:cuset for gap pdf random} and the lower bound in~\eqref{eq:inner for gap pdf random}.
Since the term $I(X_s,X_r,S_r; Y_d)$ is the same in the upper and lower bound, the gap is given by%
\footnote{
Let
a  lower bound be $\min_{\mathcal{A}}\{f_l(\mathcal{A})\}$ and 
an upper bound be $\min_{\mathcal{A}}\{f_u(\mathcal{A})\}$. 
With the definition
\[
\mathcal{A}_{u,{\rm min}} := \arg\min_{\mathcal{A}}\{f_u(\mathcal{A})\}, \
\mathcal{A}_{l,{\rm min}} := \arg\min_{\mathcal{A}}\{f_l(\mathcal{A})\},
\]
we have $f_u(\mathcal{A}_{u,{\rm min}}) \leq f_u(\mathcal{A}_{l,{\rm min}}).$
This fact implies that the gap is upper bounded as
\begin{align*}
\mathsf{GAP}
  &\leq \min_{\mathcal{A}}\{f_u(\mathcal{A})\} - \min_{\mathcal{A}}\{f_l(\mathcal{A})\} =f_u(\mathcal{A}_{u,{\rm min}}) - f_l(\mathcal{A}_{l,{\rm min}})
\\&\leq f_u(\mathcal{A}_{l,{\rm min}}) - f_l(\mathcal{A}_{l,{\rm min}})\leq \max_{\mathcal{A}}\{f_u(\mathcal{A}) - f_l(\mathcal{A})\}.
\end{align*}
}
\begin{align*}
\mathsf{GAP} 
\leq& I(X_s;Y_r,Y_d|X_r,S_r) - I(U; Y_r|X_r,S_r) 
\\&- I(X_s; Y_d|X_r,S_r,U)
\end{align*}
Next we consider two different choices for $U$:
\begin{itemize}
\item
For $C\leq S$ we choose $U=X_r$ and
\begin{align*}
&\mathsf{GAP} 
  \leq I(X_s;Y_r,Y_d|X_r,S_r) - I(X_s; Y_d|X_r,S_r)
\\&= I(X_s;Y_r|X_r,S_r,Y_d)
\\&=\!\mathbb{P}[S_r\!=\!0]I(X_s;\!\sqrt{C}X_s\!+\!Z_r|X_r,S_r\!=\!0,\!\sqrt{S}X_s\!+\!Z_d)
\\&+\mathbb{P}[S_r=1]I(X_s;            Z_r|X_r,S_r=1, \ \sqrt{S}X_s+Z_d)
\\&\leq \mathbb{P}[S_r=0]\ I(X_s;\sqrt{C}X_s+Z_r| \sqrt{S}X_s+Z_d)
\\&\leq \mathbb{P}[S_r=0]\ \log(1+C/(1+S))
\\&\leq \log(1+S/(1+S))
\\&\leq \log(2).
\end{align*}

\item
For $C> S$ we choose $U=X_rS_r+X_s(1-S_r)$ and
\begin{align*}
&\mathsf{GAP} 
 \leq I(X_s;Y_r,Y_d|X_r,S_r) 
 \\& - I(X_rS_r+X_s(1-S_r); Y_r|X_r,S_r) 
  \\&- I(X_s; Y_d|X_r,S_r,\ X_rS_r+X_s(1-S_r))
\\&=\mathbb{P}[S_r=0]\Big(I(X_s;Y_r,Y_d|X_r,S_r=0)
\\& - I(X_s; Y_r|X_r,S_r=0)\Big) 
\\&+\mathbb{P}[S_r=1]\Big(I(X_s;Y_r,Y_d|X_r,S_r=1)
\\&- I(X_s; Y_d|X_r,S_r=1)\Big) 
\\&=\mathbb{P}[S_r=0]\ I(X_s;Y_d|X_r,S_r=0,Y_r)
\\&+\mathbb{P}[S_r=1]\ I(X_s;Y_r|X_r,S_r=1,Y_d)
\\&=\!\mathbb{P}[S_r\!=\!0] I(X_s;\sqrt{S}X_s\!+\!\!Z_d|X_r,S_r\!=\!0,\!\sqrt{C}X_s\!+\!Z_r)
\\&+\mathbb{P}[S_r=1]\ I(X_s;            Z_r|X_r,S_r=1, \ \sqrt{S}X_s+Z_d)
\\&\leq \mathbb{P}[S_r=0]\ I(X_s;\sqrt{S}X_s+Z_d| \sqrt{C}X_s+Z_r)
\\&\leq \mathbb{P}[S_r=0]\ \log(1+S/(1+C))
\\&\leq \log(1+C/(1+C))
\\&\leq \log(2).
\end{align*}

\end{itemize}
This concludes the proof.
\end{IEEEproof}

\section{Proof of Proposition~\ref{prop:gap pdf det}}\label{app:prop:gap pdf det}
Consider the upper bound in~\eqref{eq:cuset for gap pdf fixed and numerical} and the lower bound in~\eqref{eq:inner for gap pdf fixed and numerical}.
Recall that $I_{1}=I_{5}$ $I_{2}=I_{6}$ $I_{3}\geq I_{7}$ $I_{4}=I_{8}$ and therefore
\begin{IEEEproof}
\begin{align*}
  \mathsf{GAP}
  &\leq \max\Big\{
  \mathcal{H}(\gamma) + \gamma I_{1}  +(1-\gamma)I_{2} - \gamma I_{5} - (1-\gamma)I_{6},
  \\&\qquad
  \gamma I_{3} +(1-\gamma)I_{4} - \gamma I_{7} - (1-\gamma)I_{8}
  \Big\}
\\&=\max\Big\{\mathcal{H}(\gamma), \gamma(I_{3}-I_{7}) \Big\}
\\&\leq \max\Big\{\log(2), \log\left(\frac{1+C P_{s,0}+S P_{s,0}}{1+\max\{C,S\}P_{s,0}}\right)\Big\}
\\&\leq \max\Big\{\log(2), \log\left(\frac{1+2\max\{C,S\}P_{s,0}}{1+\max\{C,S\}P_{s,0}}\right)\Big\}
\\&\leq \max\Big\{\log(2), \log(2)\Big\}
=\log(2) = 1~{\rm bit}.
\end{align*}
This concludes the proof.
\end{IEEEproof}

\section{Proof of Proposition~\ref{prop:gap lda det}}\label{app:prop:gap lad det}
\begin{IEEEproof}
Consider the upper bound in~\eqref{eq:cuset for gap analytical} and the lower bound in~\eqref{eq:lda inner bound}. 
We distinguish two cases:
\begin{itemize}

\item Case 1: $S > C$. In this case $r^{\rm(LDA-HD)}=\log(1+S)$. The gap is
\begin{align*}
  \mathsf{GAP} &\leq r^{\rm(CS-HD)} 
  - r^{\rm(LDA-HD)} 
\\& \leq 2\log(2) + \log \left (1+S \right )
  \frac{(b_1-1) (b_2-1)}{(b_1-1) + (b_2-1)}
  \\& \leq 2 \log(2) + \log \left( 1+S \right) \left( b_2-1 \right)
  \\& = 2 \log(2) + \log \left( 1+\frac{C}{1+S} \right) 
  \\& \leq 2 \log(2) + \log \left( 1+\frac{S}{1+S} \right)
  \\& \leq 3 \log(2) = 3 \ \rm{bits}.
\end{align*}

\item Case 2: $S \leq C$.
First,  by noticing that $\log \left(1+(\sqrt{I}+\sqrt{S})^2\right) \leq \log \left(1+I+S\right)+\log(2)$, we further upper bound the expression in~\eqref{eq:cuset for gap analytical} as
\begin{align*}
&r^{\rm(CS-HD)}
 \leq 2\log(2)+ \log \left (1+S \right ) 
\\&+ \frac{\left(\log\left(1+\frac{I}{1+S}\right) +\log(2)\right) \log\left(1+\frac{C}{1+S}\right)}{\log\left(1+\frac{I}{1+S}\right)+\log(2)+\log\left(1+\frac{C}{1+S}\right)}.
\end{align*}
Next we further lower bound $r^{\rm(LDA-HD)}$ in~\eqref{eq:lda inner bound} as
\begin{align*}
r^{\rm(LDA-HD)}
& \geq  \log \left (1+S \right ) 
\\&+ \frac{\log\left(1+\frac{I}{1+S}\right)  \left(\log\left(1+\frac{C}{1+S}\right)\!\!-\!\!\log(2)\right)}{\log\left(1+\frac{I}{1+S}\right)+\log\left(1+\frac{C}{1+S}\right)},
\end{align*}
Hence, with $x=\log\left(1+\frac{I}{1+S}\right), y=\log\left(1+\frac{C}{1+S}\right)$, we have
\begin{align*}
  \mathsf{GAP} &\leq r^{\rm(CS-HD)} 
                   - r^{\rm(LDA-HD)} 
\\&\leq 2
+ \frac{(x+1)y}{x+1+y} - \frac{x(y-1)}{x+y} 
\\&=2+\frac{x^2+y^2+xy+x}{x^2+y^2+2xy+x+y} 
\\&\leq 3~\rm{bits}.
\end{align*}
\end{itemize}
This concludes the proof.
\end{IEEEproof}

\section{Proof of Proposition~\ref{prop:gap nnc det}}\label{app:prop:gap nnc det}

\begin{IEEEproof}
With NNC we have
\begin{align*}
  &\mathsf{GAP}
  \leq \max\Big\{
  \mathcal{H}(\gamma) + \gamma I_{1}  +(1-\gamma)I_{2} - \gamma I_{9} - (1-\gamma)I_{10},
  \\&\qquad
  \gamma I_{3} +(1-\gamma)I_{4} - \gamma I_{11} - (1-\gamma)I_{12}
  \Big\}
  \\& = \max\Big\{ \mathcal{H}(\gamma) 
   + \gamma     \log \left(1+S P_{s,0}\right)  +\gamma \log \left(1+\frac{1}{\sigma_{0}^2} \right)
   \\& \qquad+ (1-\gamma) \log \left(1+(\sqrt{S P_{s,1}}+\sqrt{I P_{r,1}})^2\right)
\\& \qquad - \gamma     \log\left(1+S P_{s,0}\right) 
   - (1-\gamma) \log\left(1+S P_{s,1}+I P_{r,1} \right),
\\& \qquad \gamma     \log\left(1+(C +S) P_{s,0}\right) 
   + (1-\gamma) \log\left(1+S P_{s,1}\right)
\\& \qquad - \gamma     \log\left(1\!+\!S P_{s,0}\!+\!\frac{C P_{s,0}}{1\!+\!\sigma_{0}^2} \right)
   \!-\! (1\!-\!\gamma) \log\left( 1\!+\! S P_{s,1} \right)\Big\}
\\&\leq \max \left \{ \mathcal{H}(\gamma) + (1-\gamma)\log(2)+\gamma \log \left(1+\frac{1}{\sigma_{0}^2} \right), \right.
\\& \qquad \left. \gamma \log\left(1+\frac{\frac{\sigma_{0}^2}{1+\sigma_{0}^2}C P_{s,0}}{1+S P_{s,0}+\frac{1}{1+\sigma_{0}^2}C P_{s,0}} \right) \right \}
\\& \leq \max \left \{ \mathcal{H}(\gamma) + (1-\gamma)\log(2)+\gamma \log \left(1+\frac{1}{\sigma_{0}^2} \right), \right.
\\& \left. \qquad \qquad \gamma \log\left(1+ \sigma_{0}^2 \right) \right \}
\\& \leq 1.6081 \ \rm{bits}.
\end{align*}
where for $\sigma_{0}^2$ we chose the value
\begin{align*}
\sigma_{0}^2 = \exp{\frac{\mathcal{H}(\gamma)+(1-\gamma)\log(2)}{\gamma}}
\end{align*}
by equating the two arguments of the $\max$ (this is so because $\mathcal{H}(\gamma) + (1-\gamma)\log(2)+\gamma \log \left(1+\frac{1}{\sigma_{0}^2}\right)$ is decreasing in $\sigma_{0}^2$, while $\log\left(1+ \sigma_{0}^2 \right)$ is increasing in $\sigma_{0}^2$). 
Numerically one can find that with the chosen $\sigma_{0}^2$ the maximum over $\gamma\in[0,1]$ is 1.6081 for $\gamma=0.3855$.

Note that by choosing $\sigma_{0}^2=1$ the gap would be upper bounded by 2~bits.
\end{IEEEproof}


\section{Proof of Theorem~\ref{prop:capacity of the deterministic LDA}}
\label{app:exact capacity of LDA HD RC}
\begin{IEEEproof}
The capacity of the HD channel in~\eqref{eq:chmodel GHDRC LDA} 
is upper bounded by the capacity of the FD version of the same channel,
which is given by
\begin{align}
C^{\rm(FD)}
  &=\max_{P_{X_s,X_r}} \min \Big\{I(X_s,X_r;Y_d),I(X_s;Y_r,Y_d|X_r)\Big\} \nonumber
\\&=\max_{P_{X_s,X_r}} \min \Big\{H(Y_d),H(Y_r,Y_d|X_r)\Big\} \nonumber
\\&\stackrel{\rm (a)}{=} \min\Big\{ 
    \max\{\beta_{\rm sd},\beta_{\rm rd}\}, 
    \max\{\beta_{\rm sd},\beta_{\rm sr}\}  
\Big\} \nonumber
\\&=\beta_{\rm sd} +\min\{[\beta_{\rm rd}-\beta_{\rm sd}]^+,[\beta_{\rm sr}-\beta_{\rm sd}]^+\}.
\label{eq:exact capacity of LDA FD RC} 
\end{align}
where the equality in (a) is in general an upper bound but for this channel model is achieved with equality with i.i.d. Bernulli$(1/2)$ input bits. 

For the capacity of the HD channel we distinguish two cases:
\begin{itemize}

\item Regime 1: $\beta_{\rm rd} \leq \beta_{\rm sd}$ or $\beta_{\rm sr} \leq \beta_{\rm sd}$.

In this regime, $C^{\rm(HD)} \leq C^{\rm(FD)}=\beta_{\rm sd}$. Since the rate $C^{\rm(HD)}=\beta_{\rm sd}$ can be achieved by silencing the relay and using i.i.d. Bernulli$(1/2)$ input bits for the source, we conclude that $C^{\rm(HD)} = \beta_{\rm sd}$ in this regime.

\item Regime 2: $\beta_{\rm rd} > \beta_{\rm sd}$ and $\beta_{\rm sr} > \beta_{\rm sd}$.

We start by writing $Y_d=[Y_{d,u},Y_{d,l}]$, where 
\begin{itemize}

\item
$Y_{d,l}$ contains the lower $\beta_{\rm sd}$ bits of $Y_d$. These bits are a combination of the bits of $X_s$ and the lower bits of $X_r$. The lower bits of $X_r$ are indicated as $X_{r,l}$. With reference to Fig~\ref{fig:HDRCachlindetch relaysends}, $Y_{d,l}$ corresponds to the portion of $Y_{d}$ containing the ``orange bits'' labeled $b_1[2]$.

\item
$Y_{d,u}$ contains the upper $\beta_{\rm rd}-\beta_{\rm sd}$ bits of $Y_d$. These bits only depend on the upper bits of $X_r$. The upper bits of $X_r$ are indicated as $X_{r,u}$. With reference to Fig~\ref{fig:HDRCachlindetch relaysends}, $Y_{d,u}$ corresponds to the portion of $Y_{d}$ containing the ``green bits'' labeled $a$.

\end{itemize}

\begin{figure*}
\begin{align}
\max\{R\}
& \leq 
 I(X_{1},[X_{\mathcal{A}},S_{\mathcal{A}}]; Y_{\mathcal{A}^c},Y_{K} |[X_{\mathcal{A}^c},S_{\mathcal{A}^c}], \ S_{1}=1, S_{K}=0) \nonumber
\\& \stackrel{\rm{(a)}}{\leq} \nonumber
 H(S_{\mathcal{A}}|S_{\mathcal{A}^c}, \ S_{1}=1, S_{K}=0) +I(X_{1},X_{\mathcal{A}}; Y_{\mathcal{A}^c},Y_{K}|X_{\mathcal{A}^c},S_{[2:K-1]}, \ S_{1}=1, S_{K}=0)
%
\\& \stackrel{\rm{(b)}}{\leq}  \nonumber
  \sum_{s=0}^{2^{K-2}-1} H(S_{\mathcal{A}})
+ \sum_{s=0}^{2^{K-2}-1}\lambda_{s} \ \log\left|\mathbf{I}_{|\mathcal{A}^c|+1} + \mathbf{H}_{\mathcal{A},s} \mathbf{K}_{\{1\}\cup\mathcal{A},s} \mathbf{H}_{\mathcal{A},s}^H\right|
\\&\stackrel{\rm{(c)}}{\leq} \nonumber
|\mathcal{A}|\log(2)
+ \sum_{s=0}^{2^{K-2}-1}\lambda_{s} {\rm Rank}[\mathbf{H}_{\mathcal{A},s}]\log\left(\max\{1,{\rm Trace}[\mathbf{K}_{\{1\}\cup\mathcal{A},s}]\}\right)
\\&\qquad \nonumber
+ \sum_{s=0}^{2^{K-2}-1}\lambda_{s} \ \log\left|\mathbf{I}_{|\mathcal{A}^c|+1} + \mathbf{H}_{\mathcal{A},s}\mathbf{H}_{\mathcal{A},s}^H\right| 
\\&\stackrel{\rm{(d)}}{\leq} \nonumber
|\mathcal{A}|\log(2)
+  \min(1+|\mathcal{A}|, 1+|\mathcal{A}^c|) \ 
\log\left(\max\left\{\sum_{s=0}^{2^{K-2}-1}\lambda_{s}, \sum_{s=0}^{2^{K-2}-1}\lambda_{s} \   {\rm Trace}[\mathbf{K}_{\{1\}\cup\mathcal{A},s}]\right\}\right)
\\&\qquad \nonumber
+ \sum_{s=0}^{2^{K-2}-1}\lambda_{s} \ \log\left|\mathbf{I}_{|\mathcal{A}^c|+1} + \mathbf{H}_{\mathcal{A},s}\mathbf{H}_{\mathcal{A},s}^H\right|
\\&\stackrel{\rm{(e)}}{\leq}
|\mathcal{A}|\log(2)
+\min(1+|\mathcal{A}|, 1+|\mathcal{A}^c|)  \log\left(1+|\mathcal{A}|\right)
+ \sum_{s=0}^{2^{K-2}-1}\lambda_{s} \ \log\left|\mathbf{I}_{|\mathcal{A}^c|+1} + \mathbf{H}_{\mathcal{A},s}\mathbf{H}_{\mathcal{A},s}^H\right|
\label{eq:eqnewmulti}
\end{align}
\end{figure*}

We have
\begin{align*}
H(Y_d)
  &=H(Y_{d,u},Y_{d,l})
\\&\leq H(Y_{d,u})+H(Y_{d,l})
\\&\leq H(Y_{d,u})+\beta_{\rm sd},
\end{align*}
since $Y_{d,l}$ contains $\beta_{\rm sd}$ bits and where $H(Y_{d,u})$ is computed from the distribution
\begin{align*}
\mathbb{P}[Y_{d,u}=y]
  &=\mathbb{P}[S_r=0]\mathbb{P}[Y_{d,u}=y|S_r=0]
   \\& +\mathbb{P}[S_r=1]\mathbb{P}[Y_{d,u}=y|S_r=1]
\\&=\gamma \delta[y]
   +(1-\gamma)\mathbb{P}[X_{r,u}=y|S_r=1]
\end{align*}
for $y\in[0:L-1]$, $L:=2^{\beta_{\rm rd}-\beta_{\rm sd}}>1$, where $\delta[y]=1$ if $y=0$ and zero otherwise, and where $\gamma := \mathbb{P}[S_r=0]$.
Let $\mathbb{P}[X_{r,u}=y|S_r=1] = p_y\in[0,1] : \sum_{y}p_y=1$.
Then
\begin{align*}
&H(Y_{d,u})
 \\&=    H\Big(\big[\gamma+(1-\gamma)p_0, (1-\gamma)p_1, \ldots, (1-\gamma)p_{L-1}\big]\Big)
\\&\leq H\Big(\big[\gamma\!+\!(1\!-\!\gamma)p_0, (1\!-\!\gamma)\frac{1-p_0}{L-1}, \ldots, (1\!-\!\gamma)\frac{1-p_0}{L-1}\big]\Big)
\\&=(1-\theta)\log\frac{1}{1-\theta}
+\theta \log\frac{L-1}{\theta}|_{\theta := (1-\gamma)(1-p_0) \in[0,1-\gamma]}.
\end{align*}
The upper bound on the entropy $H(Y_{d,u})$ is maximized by 
\begin{align*}
\theta^* = 1-\max\{1/L,\gamma\} 
\Longleftrightarrow
p_0^* = \frac{[1/L-\gamma]^+}{1-\gamma}. 
\end{align*}
Therefore we have
\begin{align*}
&C^{\rm(HD)}
  \\&=\!\!\!\! \max_{P_{X_s,X_r,S_r}}\!\!\!\! \min \Big\{I(X_s,X_r,S_r;Y_d),I(X_s;Y_r,Y_d|X_r,S_r)\Big\} \nonumber
\\&= \max_{P_{X_s,X_r,S_r}} \min \Big\{H(Y_d),H(Y_r,Y_d|X_r,S_r)\Big\} \nonumber
\\&\leq \beta_{\rm sd} + 
  \max_{\gamma\in[0,1]}\min\Big\{ 
       (1-\theta^*)\log\frac{1}{1-\theta^*} +\theta^* \log\frac{L-1}{\theta^*},
 \\& \qquad \qquad \qquad \qquad      \gamma [\beta_{\rm sr}-\beta_{\rm sd}]^+
\Big\},
\end{align*}
since
\begin{align*}
&H(Y_r,Y_d|X_r,S_r)
\\&= \mathbb{P}[S_r=0]H(Y_r,Y_d|X_r,S_r=0)
\\& + \mathbb{P}[S_r=1]H(Y_r,Y_d|X_r,S_r=1)
\\&\leq 
   \gamma    \max\{\beta_{\rm sr},\beta_{\rm sd}\}
 + (1-\gamma)\beta_{\rm sr}.
\end{align*}

In order to show the achievability of~\eqref{eq:exact capacity of LDA HD RC} consider the following input: the state $S_r$ is  Bernulli$(1-\gamma)$ independent of everything else and
$X_s$ and $X_r$ are independent.
The source uses i.i.d. Bernulli$(1/2)$ bits.
The relay  uses i.i.d. Bernulli$(0)$ bits for $X_{r,l}$
and $\mathbb{P}[X_{r,u}=y] = p_0^*$ if $y=0$ and $\mathbb{P}[X_{r,u}=y] = (1-p_0^*)/(L-1)$ otherwise, i.e., the components of $X_{r,u}$ are not independent.
 Notice that the distribution of $X_{r,u}$ in state $S_r=0$ is irrelevant because its contribution at the destination is zero anyway, so we can assume that the input distribution for $X_{r}$ is independent of the state $S_r$.

\end{itemize}
\end{IEEEproof}

\begin{figure*}
\begin{align}
\max\{R\}
&\geq 
I(X_{1},X_{\mathcal{A}}; \widehat{Y}_{\mathcal{A}^c},Y_{K}| X_{\mathcal{A}^c},X_{K},S_{[2:K-1]},S_1=1,S_K=0)
 -I(Y_{\mathcal{A}};\widehat{Y}_{\mathcal{A}}| X_{[1:K]},\widehat{Y}_{\mathcal{A}^c},Y_{K},S_{[2:K-1]},S_1=1,S_K=0) \nonumber
\\&\geq
\sum_{s=0}^{2^{K-2}-1}\lambda_{s} \ \log\left|\mathbf{I}_{|\mathcal{A}^c|+1} + \frac{1}{1+\sigma^2}\ \mathbf{H}_{\mathcal{A},s}  \mathbf{H}_{\mathcal{A},s}^H\right|
-|\mathcal{A}| \log\left(1+\frac{1}{\sigma^2}\right)\nonumber
\\&\geq
\sum_{s=0}^{2^{K-2}-1}\lambda_{s} \ \log\left|\mathbf{I}_{|\mathcal{A}^c|+1} + \mathbf{H}_{\mathcal{A},s}  \mathbf{H}_{\mathcal{A},s}^H\right|
+\sum_{s=0}^{2^{K-2}-1}\lambda_{s} \ {\rm Rank}[\mathbf{H}_{\mathcal{A},s}]\log\left(\min \left \{1,\frac{1}{1+\sigma^2} \right \}\right)
-|\mathcal{A}| \ \log\left(1+\frac{1}{\sigma^2}\right)\nonumber
\\&\geq
\sum_{s=0}^{2^{K-2}-1}\lambda_{s} \ \log\left|\mathbf{I}_{|\mathcal{A}^c|+1} + \mathbf{H}_{\mathcal{A},s}  \mathbf{H}_{\mathcal{A},s}^H\right|
 - \min( 1+|\mathcal{A}|,1+|\mathcal{A}^c|)\log(1+\sigma^2)-|\mathcal{A}| \ \log\left(1+\frac{1}{\sigma^2}\right)\nonumber
\\& { {\stackrel{\sigma^2=1}{\geq}\sum_{s=0}^{2^{K-2}-1}\lambda_{s} \ \log\left|\mathbf{I}_{|\mathcal{A}^c|+1} + \mathbf{H}_{\mathcal{A},s}  \mathbf{H}_{\mathcal{A},s}^H\right| }}
 { - \min( 1+2|\mathcal{A}|,K-1)\log(2)}
\label{eq:eqnewmultilow}
\end{align}
\end{figure*}

\section{Proof of Theorem~\ref{thm:many relays}}
\label{app:prop:gap multi relay}
\begin{IEEEproof}
\paragraph{Upper bound}
The cut-set upper bound on the capacity of the HD Gaussian relay network gives, for each $\mathcal{A}$, \eqref{eq:eqnewmulti} at the top of this page 
where the inequalities are due to the following facts:
\begin{itemize}
\item
Inequality (a): chain rule of the mutual information;
\item
Inequality (b): by considering that the discrete random variable $S_{[2:K-1]}$ has at most $2^{K-2}$ masses and by letting $\lambda_{s} := \mathbb{P}[S_{[2:K-1]}=s] \in[0,1]$ for $s\in[0:2^{K-2}-1]$ such that $\sum_{s=0}^{2^{K-2}-1}\lambda_{s} = 1$. 

Here we use the convention that ``$S_{[2:K-1]}=s$'' means that the $j$-th entry of $S_{[2:K-1]}$ is equal to the $j$-th digit in the binary expansion of the number $s$. For example: with $K=5$ and $s=4 = 1 \cdot 2^2 + 0 \cdot 2^1 + 0 \cdot 2^0$, the notation ``$S_{[2:K-1]}=s$
means $S_{2}=1,S_{3}=0,S_{4}=0$''.
$\mathbf{K}_{\mathcal{A},s}$ represents the covariance matrix of $X_{\mathcal{A}}$ conditioned on $[S_{[2:K-1]}=s, \ S_{1}=1, S_{K}=0]$ and $\mathbf{H}_{\mathcal{A},s}$ is the matrix obtained from $(\mathbf{I}-\mathbf{S}) \mathbf{H} \mathbf{S}$ by retaining the rows indexed by $\{K\}\cup\mathcal{A}^c$ and the columns indexed by $\{1\}\cup\mathcal{A}$ for ${\rm diag}[\mathbf{S}]=S_{[1:K]}$ such that $[S_{[2:K-1]}=s, \ S_{1}=1, S_{K}=0]$.

Because of the power constraint we must have $\sum_{s=0}^{2^{K-2}-1}\lambda_{s} \Big[\mathbf{K}_{[1:K],s}\Big]_{k,k} \leq 1, \ k\in[1:K]$.
\item
Inequality (c): by exploiting the following relation: $0 \preceq \mathbf{K} \preceq \lambda_{\rm max}(\mathbf{K}) \mathbf{I} \preceq {\rm Trace}[\mathbf{K}]  \mathbf{I}$. 
Moreover for $a\not=0$ and by using the eigen-decomposition $\mathbf{K} = \mathbf{U} \mathbf{\Lambda} \mathbf{U}^H \in \mathbb{C}^n$ with $[\mathbf{\Lambda}]_{11} \geq [\mathbf{\Lambda}]_{22} \ldots [\mathbf{\Lambda}]_{nn}$ the following holds
\begin{align*}
& |\mathbf{I} + |a| \ \mathbf{K}| 
= |\mathbf{I} + |a| \ \mathbf{\Lambda}| 
= \prod_{j=1,...,{\rm Rank}[\mathbf{K}]}(1+|a|\lambda_j)
\\&
\leq  \max\{1,|a|\}^{{\rm Rank}[\mathbf{K}]} \ \prod_{j=1,...,{\rm Rank}[\mathbf{K}]}(1+\lambda_j)
\\&= \max\{1,|a|\}^{{\rm Rank}[\mathbf{K}]} \ |\mathbf{I} + \mathbf{K}|. 
\end{align*} 
\item
Inequality (d): since the rank of a matrix is at most the minimum between the number of rows and columns.
\item
Inequality (e): since the entropy of a discrete random variable is at most the log of the cardinality of its support and because of the input power constraints.
\end{itemize}

\paragraph{NNC lower bound}
A lower bound to the capacity of the memoryless HD-MRC is found by adapting the NNC for the general memoryless MRC~\cite{nncLim} to the HD case. In all states we consider i.i.d. $\mathcal{N} \left( 0,1 \right)$ inputs with $Q=S_{[1:K]}$ 
and with $\widehat{Y}_k := Y_k + \widehat{Z}_k$ for $\widehat{Z}_k\sim\mathcal{N}(0,\sigma^2)$ independent of everything else. 
With this, the NNC lower bound gives, for each $\mathcal{A}$, \eqref{eq:eqnewmultilow} at the top of this page.

\paragraph{Constant Gap}
The gap between cut-set upper bound and the NNC lower bound is
\begin{align*}
\mathsf{GAP}
&\leq
\max_{|\mathcal{A}|\in[0,K-2]}\Big\{
|\mathcal{A}|\log(2)\!+\!\min( 1\!+\!2|\mathcal{A}|,K\!-\!1)\log(2)
\\& \qquad +\min(1+|\mathcal{A}|, 1+|\mathcal{A}^c|)  \log\left(1+|\mathcal{A}|\right)
\Big\}
\\& = \max_{|\mathcal{A}|\in[0,K-2]}\Big\{
\min(1+|\mathcal{A}|, 1+|\mathcal{A}^c|)  \log\left(1+|\mathcal{A}|\right)
\\&\qquad 
+\min( 1+3|\mathcal{A}|,|\mathcal{A}|+K-1)\log(2)
\Big\}
\end{align*}
\end{IEEEproof}

\begin{figure*}
\begin{align}
\mathsf{GAP} 
&\leq
\min_{\sigma^2}\max_{|\mathcal{A}|\in[0:K-2]}\Big\{
\log\left(\sum_{s=0}^{2^{K-2}-1} {\mathbf{1}}_{\{\lambda_s^\star >0\}}\right) + 2\log(1+|\mathcal{A}|)
+ 2\log(1+\sigma^2)+|\mathcal{A}|\log(1+1/\sigma^2)
\Big\}\nonumber
\\&=
\log\left(\sum_{s=0}^{2^{K-2}-1} {\mathbf{1}}_{\{\lambda_s^\star >0\}}\right) + 2\log(K-1)
+ \min_{\sigma^2}\Big\{2\log(1+\sigma^2)+(K-2)\log(1+1/\sigma^2)\Big\}\nonumber
\\&=
\log\left(\sum_{s=0}^{2^{K-2}-1} {\mathbf{1}}_{\{\lambda_s^\star >0\}}\right) 
+ 2\log(K-1)
+2\log\left(1+\frac{K-2}{2}\right)+(K-2)\log\left(1+\frac{2}{K-2}\right)\nonumber
\\&\leq
\log\left(\sum_{s=0}^{2^{K-2}-1} {\mathbf{1}}_{\{\lambda_s^\star >0\}}\right) 
+2\log\Big(K(K-1)\Big)+2\log({\rm e}/2)\nonumber
\\&\leq(K-2)\log(2)+4\log(K)+2\log({\rm e}/2).
\label{eq:eqnewgap}
\end{align}
\end{figure*}

\section{Proof of Proposition~\ref{prop:hd diamond}}
\label{app:hd diamond}
\begin{IEEEproof}
The proof of Proposition \ref{prop:hd diamond} follows directly from the proof of Theorem \ref{thm:many relays}, by taking into consideration the following facts:
\begin{itemize}
\item 
In the diamond network the channel matrix $\mathbf{H}$ has rank 2. Thus ${\rm Rank}[\mathbf{H}_{\mathcal{A},s}]=2$ both in the cut-set upper bound and in the NNC lower bound;
\item 
Instead of the upper bound $H(S_{\mathcal{A}}) \leq |\mathcal{A}| \log(2)$ we proceed as follows
\[
H(S_{\mathcal{A}}) \leq H(S_{[2:K-2]}) \leq \log\left(\sum_{s=0}^{2^{K-2}-1} {\mathbf{1}}_{\{\lambda_s >0\}}\right)
\]
where ${\mathbf{1}}_{\{\lambda_s >0\}}$ is the indicator function defined as
\begin{align*}
{\mathbf{1}}_{\{\lambda_s >0\}} = \left \{ 
\begin{array}{ll}
1 & {\text{if}} \ \lambda_s >0\\
0 & {\text{otherwise}}
\end{array}.
 \right.
\end{align*}
\end{itemize}
Let $\lambda_s^\star$ be the optimal values from the cut-set upper bound.
Then the gap is given by~\eqref{eq:eqnewgap} at the top of next page. The optimal value of the quantization noise is $\sigma^2 = K/2-1$ and it has been found by equating to zero the first derivative of $f(\sigma^2)=2\log(1+\sigma^2)+(K-2)\log(1+1/\sigma^2)$. Now to be sure that this $\sigma^2$ is the one that minimizes $f(\sigma^2)$ we need to compute the second derivative that is given by
\begin{align*}
f''(\sigma^2)=\frac{-2 \sigma^4+(K-2)(2\sigma^2+1)}{(\sigma^2+1)^2\sigma^4}.
\end{align*}
The denominator of $f''(\sigma^2)$ is always a positive number. Now we need to evaluate the numerator of $f''(\sigma^2)$ in $\sigma^2= K/2-1$ and verify that it is always positive. We obtain
\begin{align*}
&-2 \sigma^4\!+\!(K\!\!-2)(2\sigma^2\!+\!1) \!=\! -2 \left( \frac{K\!-\!2}{2} \right)^2\!+\!(K\!-\!2)(K\!-\!1)
\\&=K \left( \frac{K}{2}-1 \right) \geq 0 \ \forall \ K \geq 3.
\end{align*}
This shows that for a diamond network the optimal quantization noise is $\sigma^2 = K/2-1$ and that the gap has two components: one depends on how many non-zero $\lambda_s^\star, s\in[0:2^{K-2}-1],$ are needed to attain the cut-set upper bound (which in general is upper bounded by $2^{K-2}$), and the other is logarithmic in the number of nodes in the network.

Recently, it has been shown numerically for diamond networks with at most seven relays~\cite{Fragouli2012}, i.e., $K\leq 7$, that $\sum_{s=0}^{2^{K-2}-1} {\mathbf{1}}_{\{\lambda_s^\star >0\}} \leq K-1$. In the same work it has been conjectured that the same holds for any number of relays. If the conjectured is true then our gap would be at most $5\log(K)+2\log({\rm e}/2)$.

\end{IEEEproof}

\clearpage
\newpage

%

\begin{figure*}
\centering
\includegraphics[width=0.8\textwidth]{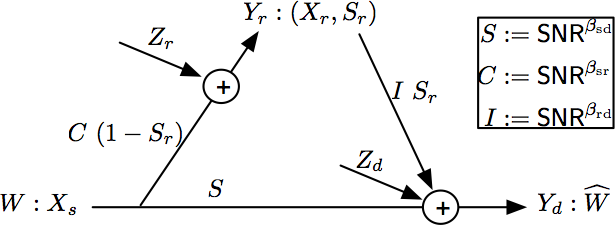}
\caption{The Half-Duplex Gaussian Relay Channel.}
\label{fig:fig1}

\subfigure[HD Phase~I ($S_r=0$).]{
\includegraphics[width=0.9\columnwidth]{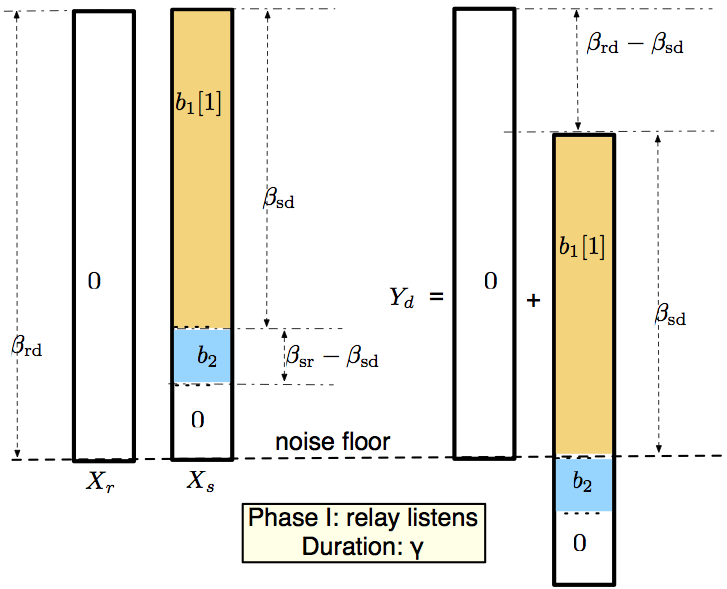}%
\label{fig:HDRCachlindetch relayreceivers}%
}
\hfill
\subfigure[HD Phase II ($S_r=1$).]{
\includegraphics[width=0.9\columnwidth]{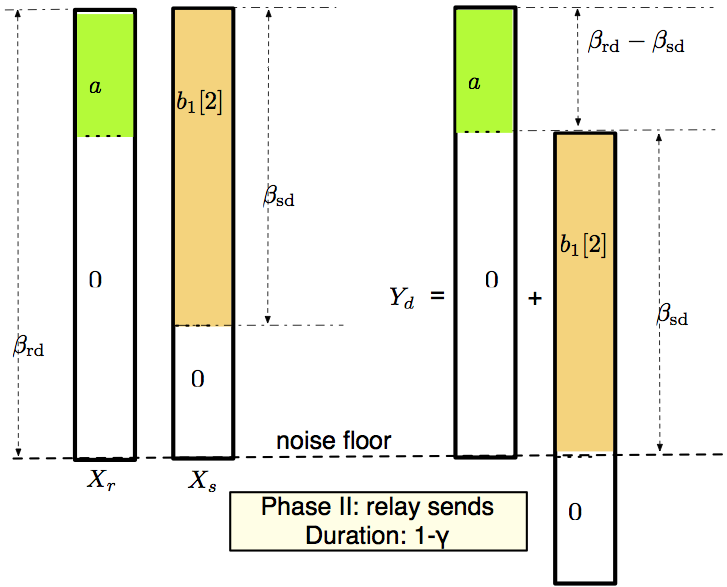}%
\label{fig:HDRCachlindetch relaysends}%
}
\hfill
\subfigure[FD.]{
\includegraphics[width=0.9\columnwidth]{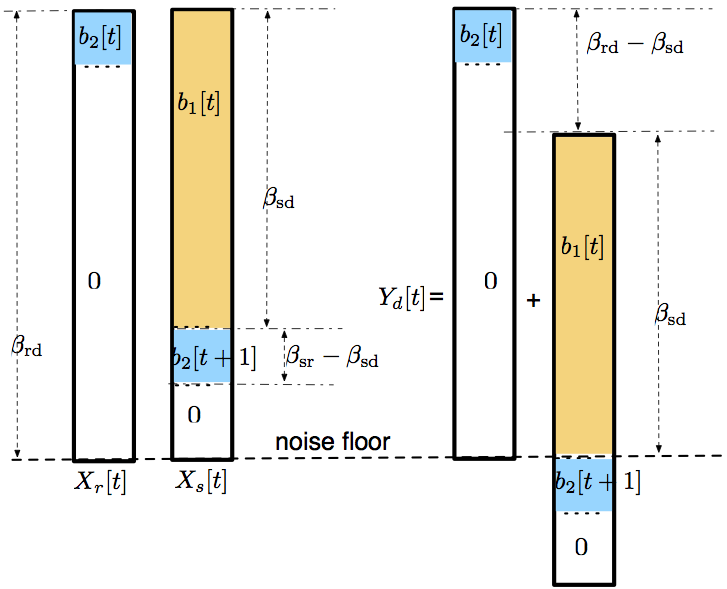}%
\label{fig:FDRCachlindetch}%
}
\caption{The gDoF optimal strategy for the linear deterministic approximation of the Gaussian noise channel at high SNR.
HD in Figs.~\ref{fig:HDRCachlindetch relayreceivers} and~\ref{fig:HDRCachlindetch relaysends} and FD in Fig.~\ref{fig:FDRCachlindetch}}
\label{fig:HDRCachlindetch}
\end{figure*}

\begin{figure*}
\centering
\includegraphics[width=0.75\textwidth]{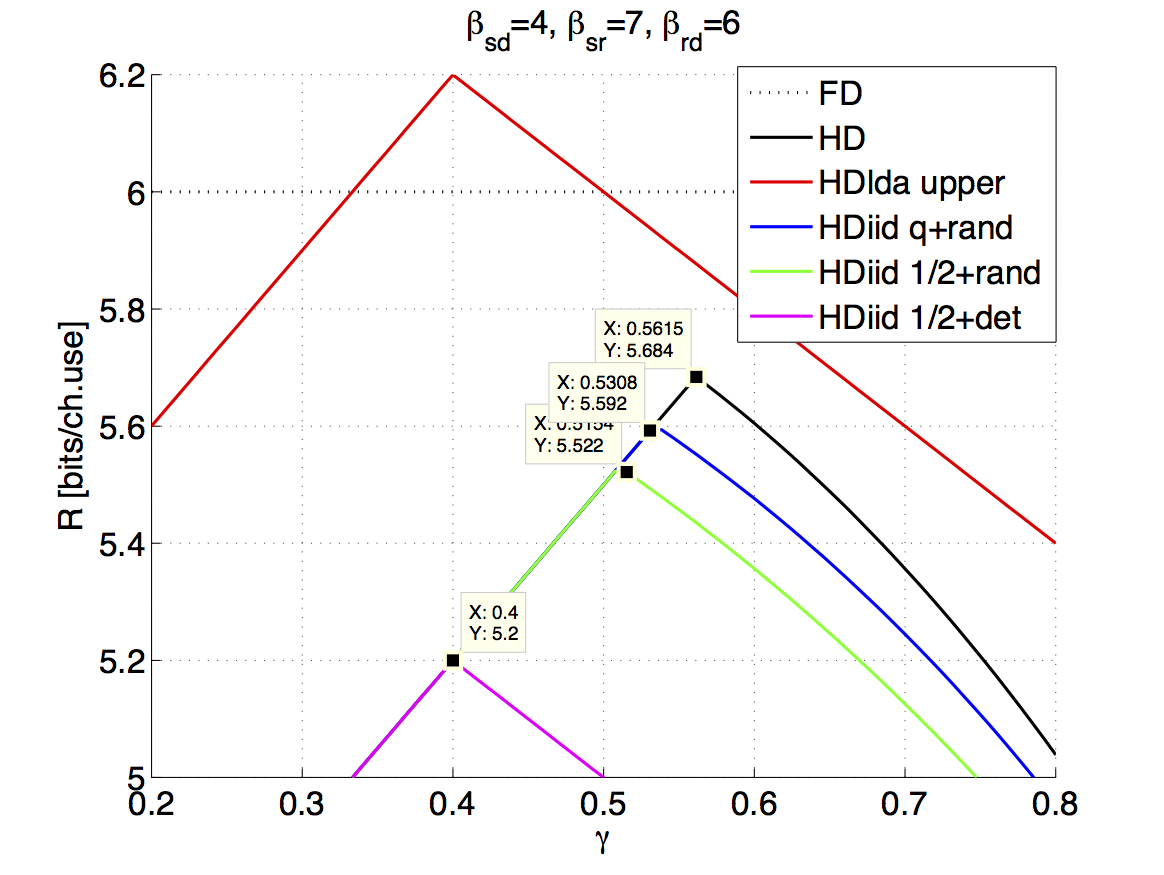}
\caption{Comparison of the capacities of the LDA for both HD and FD. Also, upper and lower bounds for the capacity of the HD channel.}
\label{fig:ldahdfig}

\centering
\includegraphics[width=0.65\textwidth]{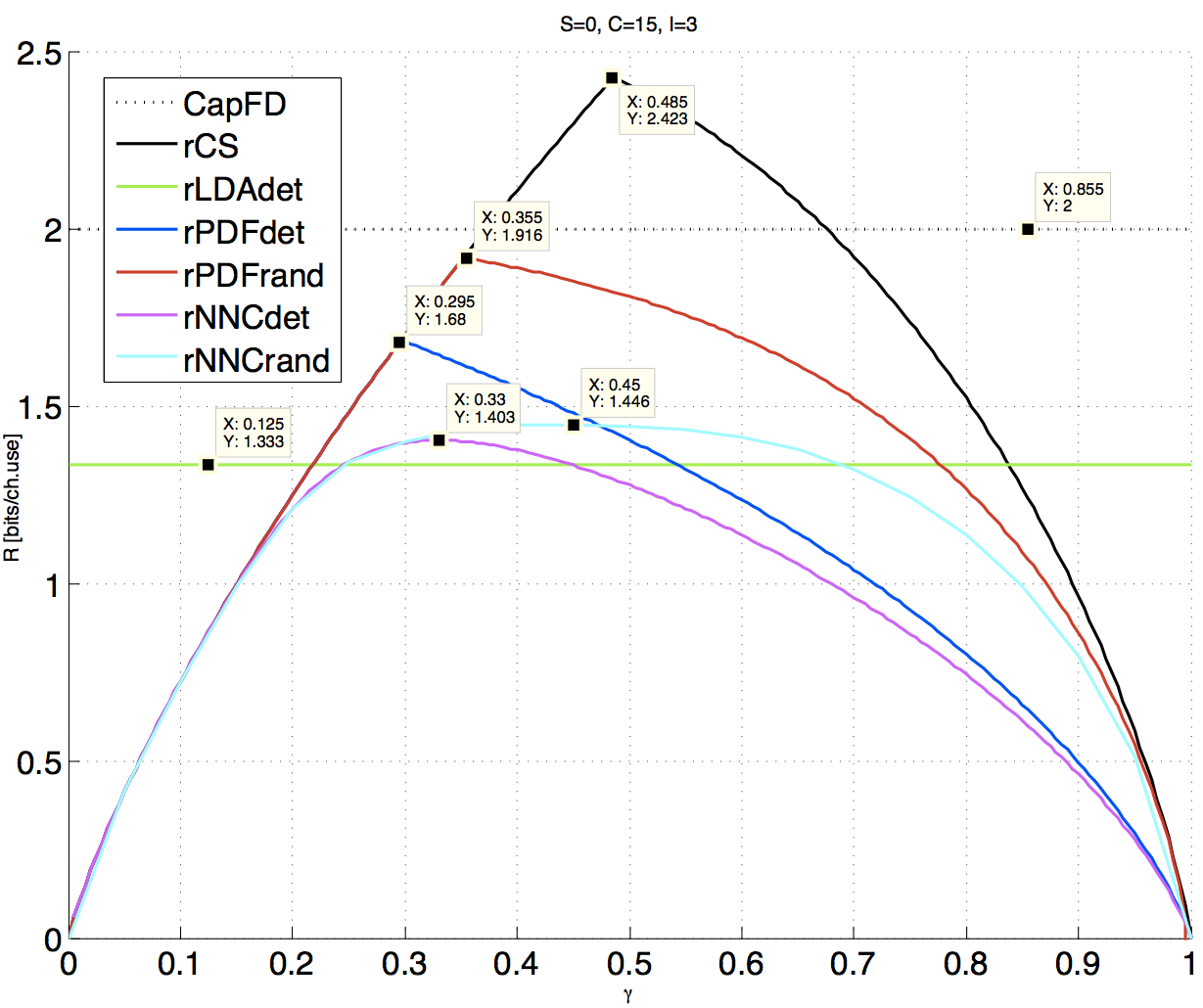}
\caption{Comparison of the rates of the G-RC without a direct link for $S=0, \ C=15, \ I=3$.}
\label{fig:diamond1relayhdfig}
\end{figure*}

\begin{figure*}
\centering
\includegraphics[width=0.65\textwidth]{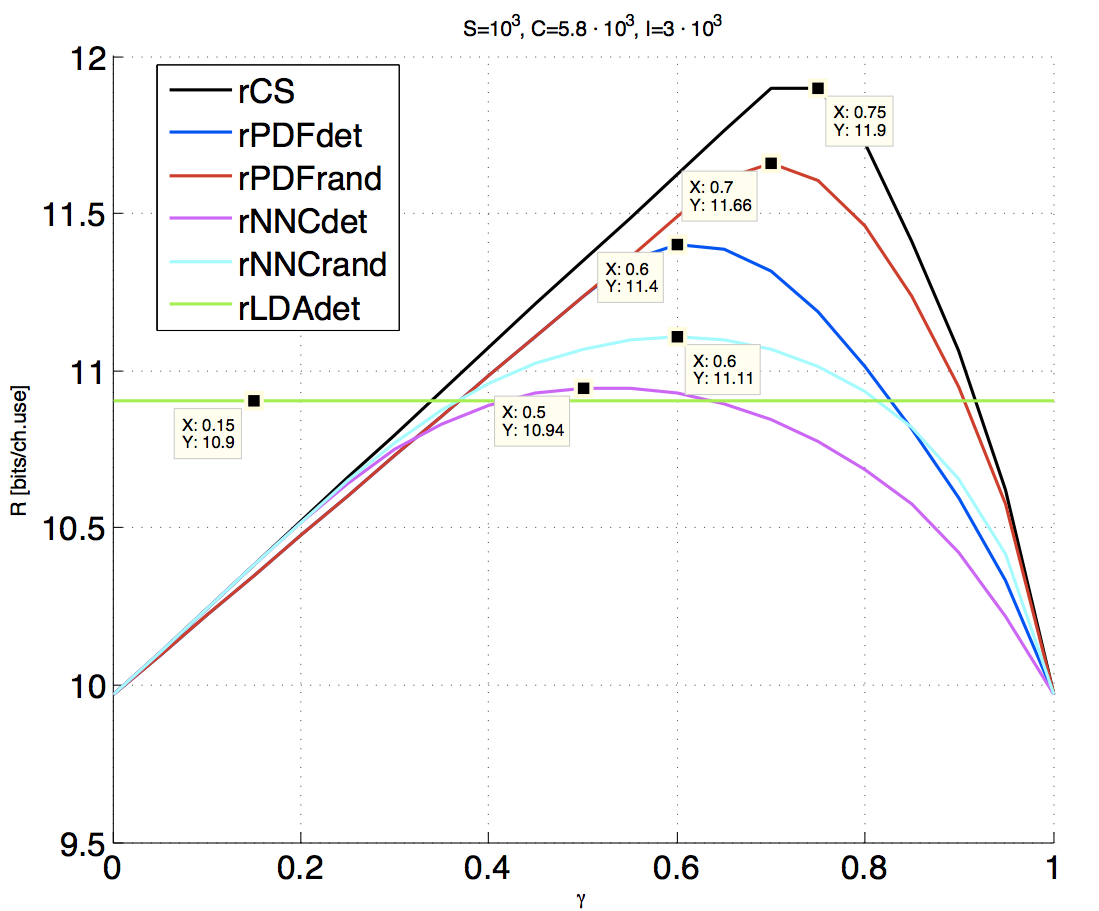}
\caption{Comparison of the rates of the G-RC for $S=30 \ \rm{dB}, \ C=37.63 \ \rm{dB}, \ I=34.77 \ \rm{dB}$.}
\label{fig:PDFbetter}

\centering
\includegraphics[width=0.65\textwidth]{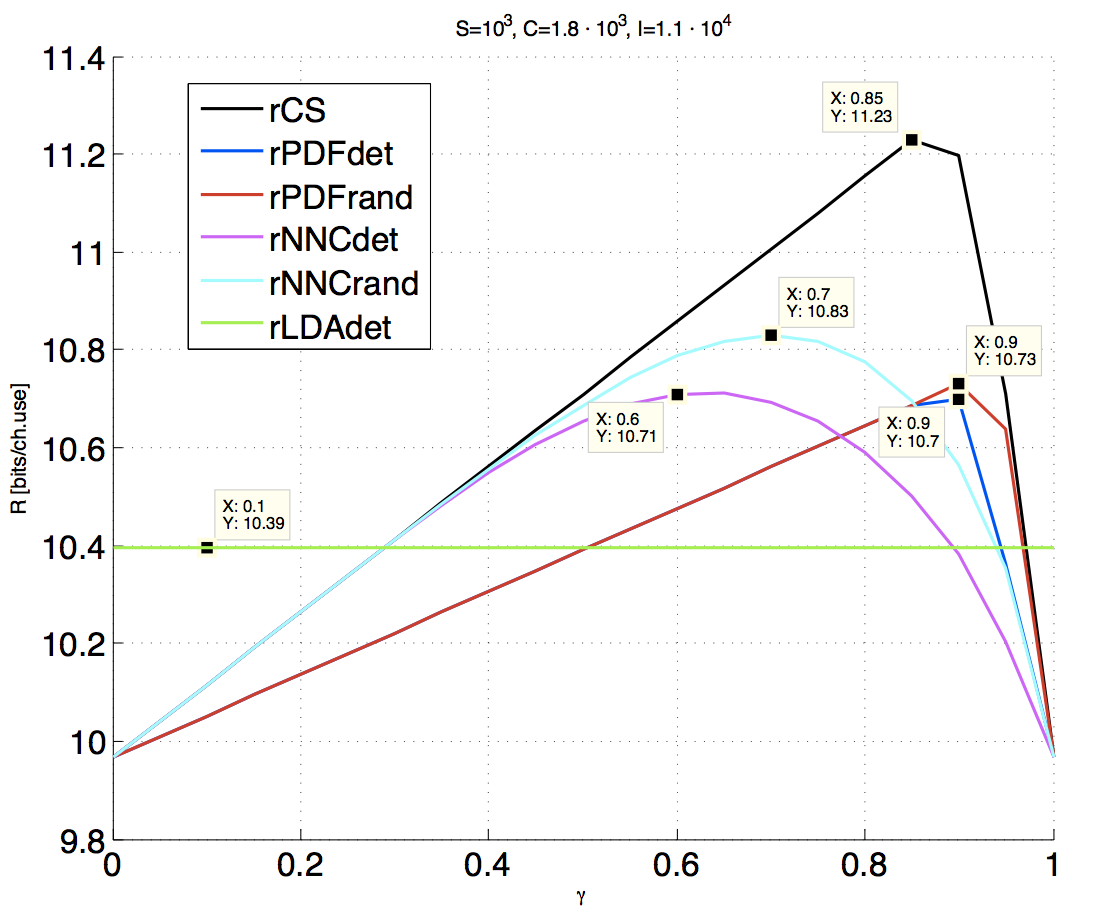}
\caption{Comparison of the rates of the G-RC for $S=30 \ \rm{dB}, \ C=32.55 \ \rm{dB}, \ I=40.41 \ \rm{dB}$.}
\label{fig:NNCbetter}
\end{figure*}

\begin{figure*}
\centering
\includegraphics[width=1.3\columnwidth]{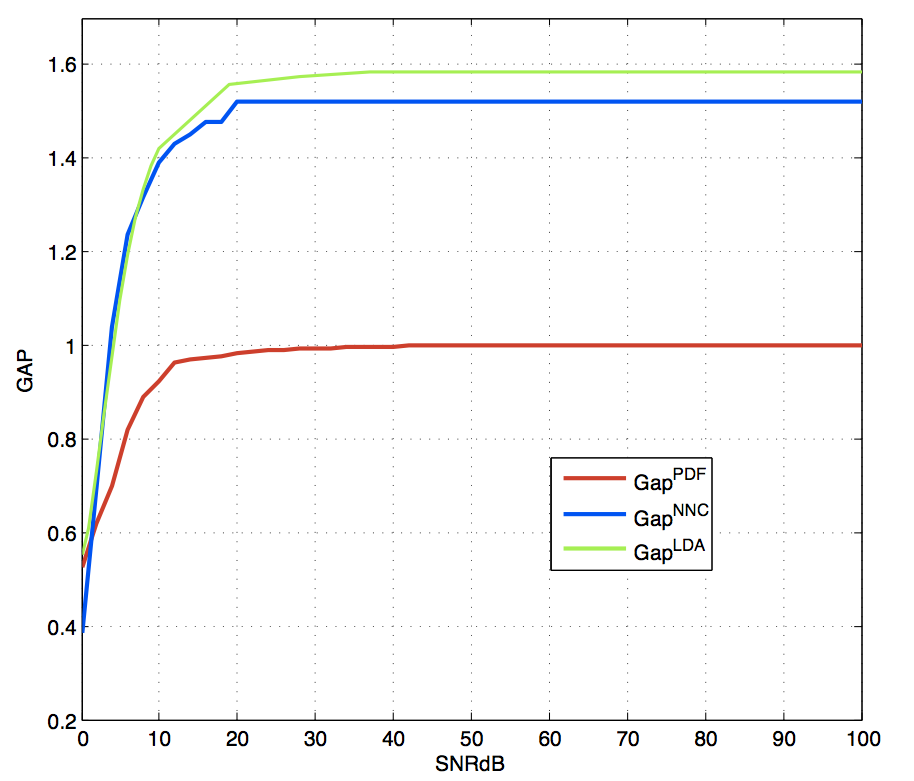}
\caption{Numerical evaluation of the maximum gap varying the SNR for $\beta_{\rm sd} =1$ and $(\beta_{\rm rd},\beta_{\rm sr})\in[0,2.4]$ with deterministic switch.}
\label{fig:actuallgapI0zero}

\centering
\includegraphics[width=1.3\columnwidth]{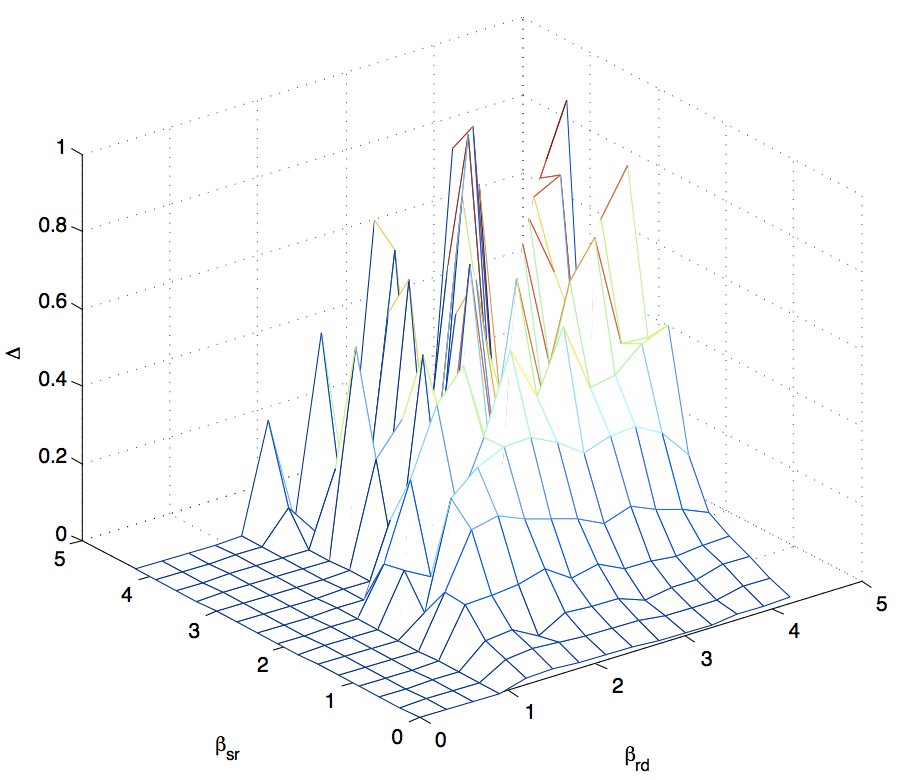}
\caption{$\Delta = r^{\rm(PDF-HD)}-r^{\rm(PDF-HD)}|_{I_0=0}$ at $\rm{SNR}=20\rm{dB}$ for $\beta_{\rm sd} =1$ as a function of $(\beta_{\rm rd},\beta_{\rm sr})\in[0,2.4]$.}
\label{fig:actuallgapPDF}
\end{figure*}

\begin{figure*}
\centering
\includegraphics[width=1.3\columnwidth]{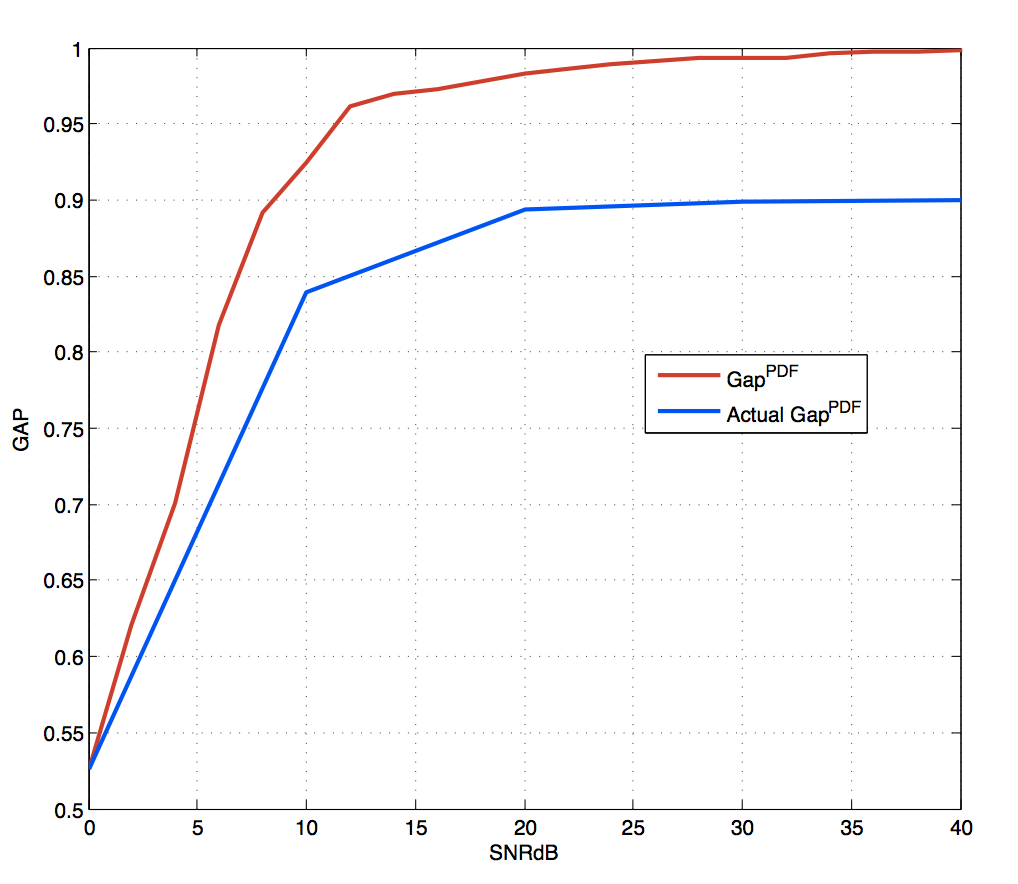}
\caption{Numerical evaluation of the maximum gap varying the SNR for $\beta_{\rm sd} =1$ and $(\beta_{\rm rd},\beta_{\rm sr})\in[1,2.4]$ with deterministic switch (red curve) and random switch (blue curve).}
\label{fig:gapPDFdetrandom}

\centering
\includegraphics[width=1.3\columnwidth]{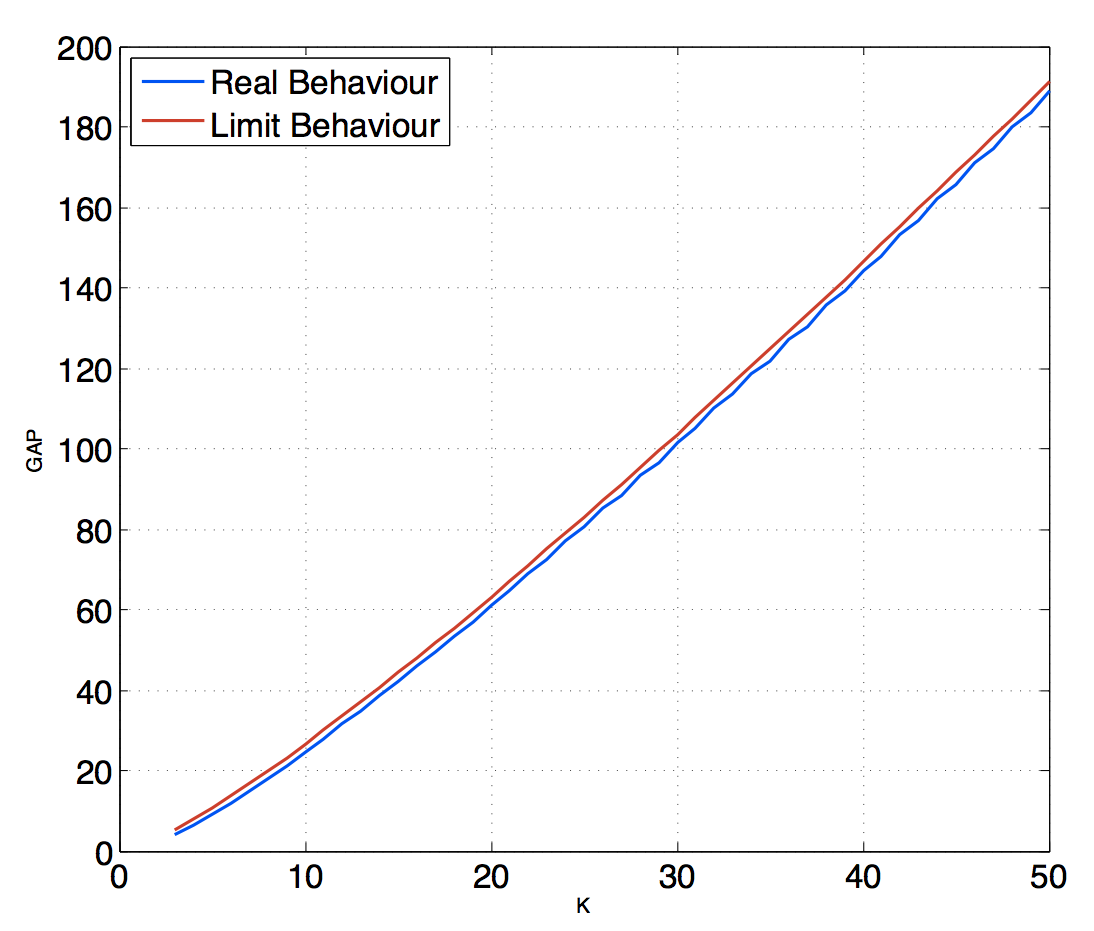}
\caption{Gap in \eqref{eq:GapMultipleRelay} (blue curve) and limit behaviour in \eqref{eq:GapMultipleRelayAsympt} (red curve) as a function of $K$ with $\sigma^2=1$.}
\label{fig:GapVaryingUsers}
\end{figure*}

\begin{figure*}
\centering
\includegraphics[width=1.5\columnwidth]{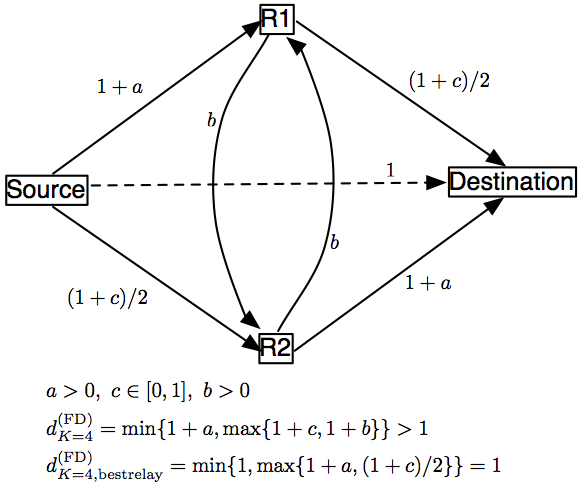}
\caption{Example of a full-duplex 2-relay network with gDoF strictly larger than the gDoF obtained by using the best relay only. The numerical value on a link represents the SNR exponent on the corresponding link.}
\label{fig:tworelaynet}
\end{figure*}

\end{document}